\documentclass[usenatbib,a4paper]{mn2e}

\usepackage[dvips]{graphicx}
\usepackage{epsfig}
\usepackage{amssymb}

\def\reff@jnl#1{{\rm#1\/}}
\def\apj{\reff@jnl{ApJ}}       
\def\apjl{\reff@jnl{ApJ}}      
\def\apjs{\reff@jnl{ApJS}}     
\def\aaps{\reff@jnl{A\&AS}}    
\def\mnras{\reff@jnl{MNRAS}}   
\def\prd{\reff@jnl{Phys.\ Rev.\ D}} 

\title{Estimating the bispectrum of the Very Small Array data}
\author[Sarah Smith et al.]
{Sarah Smith,$^1$\thanks{E-mail: sjm84@mrao.cam.ac.uk (SS);
graca@mrao.cam.ac.uk (GR); a.d.challinor@mrao.cam.ac.uk (AC)}
 Gra\c{c}a Rocha,$^{1,4,5}$\footnotemark[1]	 
Anthony Challinor,$^1$\footnotemark[1] 
Richard A. Battye,$^2$
\newauthor
 Pedro Carreira,$^2$ 
 Kieran Cleary,$^2$ 
 Rod D. Davies,$^2$ 
 Richard J. Davis,$^2$  
\newauthor 
 Clive Dickinson,$^2$
 Ricardo Genova-Santos,$^3$
 Keith Grainge,$^1$
 Carlos M. Guti{\'e}rrez,$^3$  
\newauthor
 Yaser A. Hafez,$^2$
 Michael P. Hobson,$^1$  
 Michael E. Jones,$^1$
 R\"udiger Kneissl,$^1$ 
\newauthor
 Katy Lancaster,$^1$ 
 Anthony Lasenby,$^1$  
 J. P. Leahy,$^2$
 Klaus Maisinger,$^1$  
\newauthor 
 Guy G. Pooley,$^1$ 
 Nutan Rajguru,$^1$ 
 Rafael Rebolo,$^{3,6}$ 
 Jos\'e Alberto Rubi\~no-Martin,$^3$\thanks{Present address:
 Max-Planck Institut f\"ur Astrophysik, Garching, Germany.}
\newauthor 
 Pedro Sosa Molina,$^3$  
 Richard D.E. Saunders,$^1$
 Richard S. Savage,$^1$\thanks{Present address:
 Astronomy Centre, University of Sussex.}, 
 Paul Scott,$^1$  
\newauthor 
 An\v ze Slosar,$^1$\thanks{Present address: Faculty of Mathematics
and Physics, University of Ljubljana, Slovenia.} 
 Angela C. Taylor,$^1$
 David Titterington,$^1$  
 Elizabeth Waldram$^1$
\newauthor 
 and Robert A. Watson$^2$\thanks{Present address: Instituto 
de Astrof{\'{\i}}sica de Canarias.}
\\
  $^1$Astrophysics Group, Cavendish Laboratory, University of Cambridge, 
  Madingley Road, Cambridge CB3 0HE\\
  $^2$ University of Manchester, Jodrell Bank Observatory\\
  $^3$ Instituto de Astrof{\'i}sica de Canarias, 38200 La Laguna,
  Tenerife, Spain\\
  $^4$ Department of Physics, Nuclear \& Astrophysics Laboratory,
  University of Oxford, Keble Road, Oxford OX1 3RH\\
  $^5$ Centro de Astrof\'{\i}sica da Universidade do Porto, R. das
  Estrelas s/n, 4150-762 Porto, Portugal\\
  $^6$ Consejo Superior de Investigaciones Cient{\'{\i}}ficas, Spain}

\pagerange{\pageref{firstpage}--\pageref{lastpage}}
\pubyear{2004}

\begin{document}

\maketitle

\label{firstpage}

\begin{abstract}
We estimate the bispectrum of the Very Small Array data from the compact
and extended configuration observations released in December 2002,
and compare our results to those obtained from Gaussian simulations.
There is a slight excess of large
bispectrum values for two individual fields, but this does not
appear when the fields are combined. Given our expected level of
residual point sources, we do not expect these to be the source 
of the discrepancy. Using the compact
configuration data, we put an upper limit of 5400 on the value of 
$f_{\mathrm{NL}}$, the non-linear coupling parameter, at 95 per cent confidence.
We test our bispectrum estimator using
non-Gaussian simulations with a known bispectrum, and recover
the input values.
\end{abstract}



\section{Introduction}

Currently-favoured cosmological theories predict that the primordial
fluctuations in the cosmic microwave background (CMB) obey Gaussian
statistics to a high degree. The majority of inflationary scenarios
imply a level of non-Gaussianity that is unlikely to be detectable
by any forthcoming experiment \citep{Acquaviva, Maldacena}, although
non-linear gravitational evolution may result in detectable 
non-Gaussianity from the initially almost-Gaussian fluctuations
\citep*{Bartolo}. Any convincing evidence for a departure
from primordial Gaussianity would therefore play a very significant
role in constraining theories of inflation. However, the dominant
contributions to non-Gaussianity in the CMB are expected to come from
secondary effects such as gravitational lensing, reionization, the
Sunyaev-Zel'dovich effect, and from the local Universe. These effects
are of varying significance, depending on the scale. Coupling between
lensing and the Sunyaev-Zel'dovich effect tends to dominate over 
primordial non-Gaussianity on small scales \citep{Goldberg}.
The effects of dust and gas clouds are
very dependent on the region of sky observed. The Very Small Array
(VSA) fields were carefully chosen to minimise
contamination from Galactic emission and bright radio sources,
and the signal from Galactic foregrounds is thought to be much less than 
the primordial CMB signal, with the amount of contamination smaller
on smaller scales \citep{Taylor}. The recent results from the
Wilkinson Microwave Anisotropy Probe ({\sl WMAP}; \citealt{WMAP})
have considerably tightened the constraints on primordial
non-Gaussianity, whilst detecting a non-Gaussian signal arising from
residual point sources. The non-Gaussian signal that was found in the
bispectrum of the COBE data \citep*{COBE_ng}
has not been replicated in the {\sl WMAP} data and is now believed
to be a result of systematic errors \citep{Magueijo}.
The VSA has a dedicated point source subtractor,
in order to remove all the sources above a certain minimum flux level,
so that the residual contribution from unsubtracted sources is
less than the flux sensitivity \citep{Taylor}.

Whilst we may therefore have little basis for expecting to detect primordial
non-Gaussianity in the VSA data, it is important to test the data
nevertheless, if only to validate the assumption of Gaussianity
which is made during the estimation of the power spectrum and 
its errors \citep{Hobson}.
Currently the published VSA data extend past $\ell=1400$ \citep{Grainge},
in comparison
with the current {\sl WMAP} data which do not go beyond $\ell=900$,
so we are probing the fluctuations at higher resolution.
Testing the data may also help to ascertain whether point source subtraction
has been performed satisfactorily, and to see if there is any contamination
by foreground sources. 

The VSA data have already been tested for non-Gaussianity using a variety
of statistics in the map plane, and in the visibility plane by adopting
a non-Gaussian likelihood function. The results are presented in
\citet{Savage}. Here, we use the bispectrum, the three-point
statistic in the visibility plane.

In Section \ref{sec:stats} we discuss the statistics of CMB temperature
fluctuations, and how the level of non-Gaussianity can be measured
by the bispectrum. 
We then give a brief overview of the VSA in Section \ref{sec:VSA},
particularly with reference to the point source subtraction technique,
and present an approximate expression for the measured 
visibility three-point function,
taking into account the convolution with the primary beam. An exact
expression is derived in Appendix \ref{sec:appc}.
In Section \ref{sec:calcbi} we describe our method for estimating the
bispectrum of the VSA data, which we have tested by producing
non-Gaussian simulations as described in Section \ref{sec:sims}.
We then discuss a method for comparing the results to Gaussian simulations
in Section \ref{sec:tests} and present our results. In Sections \ref{sec:point}
and \ref{sec:simp} we consider the effect of point sources on the bispectrum,
and look at the feasibility of detecting them in this way. Finally, in
Section \ref{sec:fnl} we investigate the constraints that the VSA data
are able to place on primordial non-Gaussianity. In the appendices we
discuss optimal cubic bispectrum estimators, and the effect of the
primary beam on the observed bispectrum from interferometric data.


\section{Statistics of CMB temperature fluctuations}
\label{sec:stats}

\subsection{Power spectrum}

Assuming full sky coverage, the temperature fluctuations of the CMB can be
decomposed into spherical harmonics ($Y_{\ell m}$), and hence expressed as
\begin{equation}
\frac{\Delta T}{T}\left( \bmath{\hat{n}}\right) 
=\sum _{\ell,m}a_{\ell m}Y_{\ell m}\left( \bmath{\hat{n}}\right) .
\end{equation}
Here, $T = T_0$, the mean temperature of the CMB.
Rotational invariance demands that 
$\left\langle a_{\ell_1 m_1}a^{*}_{\ell_2 m_2}\right\rangle 
=C_{\ell_1}\delta _{\ell_1 \ell_2}\delta _{m_1 m_2},$
where the brackets denote the ensemble average value. The values of the
$C_{\ell}$ 
represent the power spectrum of the CMB, which has been the major
focus of attention in CMB studies. The measured values of the $C_\ell$  will
differ slightly from the ensemble-averaged $C_{\ell}$ 
due to instrument noise and cosmic variance; if the $a_{\ell m}$ 
are each drawn independently from 
a Gaussian distribution, with mean 0 and variance $C_{\ell}$,
then the measured $C_\ell$ will be drawn from a $\chi^2$ distribution,
with an intrinsic variance equal to 
$ \frac{2 C_\ell^2}{2\ell+1} $.\footnote{In general the $ a_{\ell m}$
are complex, and the variance of the real and imaginary parts is 
$C_\ell/2$, but since they satisfy the relation
$a_{\ell m}=(-1)^{m}a_{\ell-m}^{*}$ they have $2\ell+1$ degrees of freedom.}
Therefore, particularly at low $\ell$, there will always be an uncertainty
as to the true value of the $C_\ell$, independent of the technical 
difficulties of experimental measurement.

\subsection{Higher-order statistics}

If the temperature fluctuations of the CMB are Gaussian, the power
spectrum completely describes the statistics. However, if there is
a departure from Gaussianity, higher-order statistics are needed for
a full description. In principle there is an infinite number of ways
in which the CMB could be non-Gaussian, and therefore the optimal
statistic to use depends on the type of non-Gaussianity present. If
we are looking for a particular signal we can use a statistic which
is tailored for optimal detection of that signal. This method has
been used to detect non-Gaussianity arising from point sources in
the {\sl WMAP} data \citep{WMAP}. However, if we are not seeking a specific
signature, then we require something which is fairly general. The
natural follow-on from the power spectrum is the bispectrum, which
is the three-point function given by 
\begin{equation}
B^{m_{1}m_{2}m_{3}}_{\ell_{1}\ell_{2}\ell_{3}} 
\equiv \left\langle a_{\ell_{1}m_{1}}a_{\ell_{2}m_{2}}a_{\ell_{3}m_{3}}
\right\rangle .
\end{equation}
The bispectrum gives a scale-dependent measure of skewness \citep{Santos}. 
The ensemble-average bispectrum will be zero in the case of Gaussian 
fluctuations. However, for a given realisation, it will be non-zero, 
even neglecting instrumental noise and resolution, owing to cosmic variance.

The ensemble-averaged bispectrum is constrained by the assumption
of rotational invariance in a similar way to the power spectrum. This
means that all the information is contained in the dependence of the
bispectrum on the values of $\ell$, and not in its dependence on
$m$. It can be shown that
\begin{equation}
\left\langle a_{\ell_{1}m_{1}}a_{\ell_{2}m_{2}}a_{\ell_{3}m_{3}}\right\rangle 
=\left( 
\begin{array}{ccc}\ell_{1} & \ell_{2} & \ell_{3}\\
	          m_{1} & m_{2} & m_{3}
\end{array}
\right) 
 B_{\ell_{1}\ell_{2}\ell_{3}},
\end{equation}
where the matrix represents the Wigner
3-$j$ symbol \citep{Rotenberg}. Using the relation 
\begin{equation}
\sum _{m_{1}m_{2}m_{3}}\left( 
\begin{array}{ccc} \ell_{1} & \ell_{2} & \ell_{3}\\
		   m_{1} & m_{2} & m_{3}
\end{array}
\right) ^{2}=1,
\end{equation}
we find that we can estimate the bispectrum as 
\begin{equation}
\widehat B_{\ell_{1}\ell_{2}\ell_{3}}= \!\!\!
\sum _{m_{1}m_{2}m_{3}} \! \left( 
\begin{array}{ccc} \ell_{1} & \ell_{2} & \ell_{3}\\
		   m_{1} & m_{2} & m_{3}
\end{array}
\right) a_{\ell_{1}m_{1}}a_{\ell_{2}m_{2}}a_{\ell_{3}m_{3}}.
\end{equation}

If the CMB is Gaussian, then the cosmic variance of the bispectrum
can be shown to be 
\begin{eqnarray}
\lefteqn{\left\langle \widehat B_{\ell_{1}\ell_{2}\ell_{3}}^{2}
\right\rangle} &&\\
\lefteqn{= C_{\ell_{1}} C_{\ell_{2}} C_{\ell_{3}} 
\left( 1+2\delta _{\ell_{1}\ell_{2}}\delta _{\ell_{2}\ell_{3}}
+\delta _{\ell_{1}\ell_{2}}+\delta _{\ell_{2}\ell_{3}}
+\delta _{\ell_{3}\ell_{1}}\right)}. && \nonumber
\end{eqnarray}
The variance is twice as large if two $\ell$s are the same,
or six times as large if all the $\ell$s are equal. This variance is
enhanced if we only observe a portion of the sky (sample variance). 
In real experiments the measured signal will have a contribution
arising from instrumental noise, which will be an additional source
of variance. It is usually reasonable to assume
that the noise is Gaussian, but it is conceivable that this
could contribute to a non-zero bispectrum. 

In general, we can define any $n$-point function in harmonic space in
a similar way to the power spectrum and bispectrum, and use these
as a test for non-Gaussianity \citep{Hu}. However, definite detection 
of a signal arising from true non-Gaussianity at 
higher orders is progressively more unlikely as $n$ increases. 

\subsection{Flat-sky approximation}

If we are only observing a small patch of sky then we can use the
flat-sky approximation \citep{Hu1}. Instead of decomposing the
temperature fluctuation
into spherical harmonics, we instead use Fourier modes, so that 
the temperature fluctuations can now be expressed as 
\begin{equation}
\frac{\Delta T}{T}\left( \bmath{\hat{x}}\right) 
=\int a(\bmath{u})
 \exp (2\pi i\bmath{u\cdot\hat{x}})\mathrm{d^{2}}\bmath{u}.
\end{equation}
Translational and rotational invariance demand that 
\[ \left\langle a(\bmath{u})a(\bmath{u}')^{*}\right\rangle 
=C(\left| \bmath{u}\right| )\delta ^{2}(\bmath{u}-\bmath{u}'),\]
and there is a straightforward correspondence with the full sky power 
spectrum at large $\ell$ :
$ C(\left| \bmath{u}\right| )
 \approx C_{\ell}\mid _{l=2\pi \left| \bmath{u}\right| }$. Similarly, 
assuming rotational and parity invariance,
\begin{equation}
\left\langle a(\bmath{u}_{1})a(\bmath{u}_{2})
	     a(\bmath{u}_{3})\right\rangle 
=B(\ell_{1},\ell_{2},\ell_{3})\delta ^{2}(\bmath{u}_{1}
      +\bmath{u}_{2}+\bmath{u}_{3}),
\label{eq:flat_bi}
\end{equation}
with the correspondence $\ell_{i}=2\pi \left| \bmath{u}_{i}\right| $. 
Isotropy demands that the bispectrum is
zero unless the three $u$-vectors sum to zero, so the
bispectrum is specified by just two vectors, hence its name.
Isotropy and parity invariance demand that the 
permutation-symmetric $B(\ell_1,\ell_2,\ell_3)$ depends only on the lengths
of $\bmath{u}_1$,$\bmath{u}_2$ and $\bmath{u}_3$.  At large
$\ell$ we can write $ B(\ell_{1},\ell_{2},\ell_{3})\approx 
b_{\ell_{1}\ell_{2}\ell_{3}} $,
where $b_{\ell_{1}\ell_{2}\ell_{3}}$ is the reduced bispectrum
\citep{Komatsu_fnl}: 
\[
B_{\ell_{1}\ell_{2}\ell_{3}}
=\sqrt{\frac{(2\ell_{1}\!+\!1)(2\ell_{2}\!+\!1)(2\ell_{3}\!+\!1)}{4\pi }}
\left( 
\begin{array}{ccc}
\ell_{1} & \ell_{2} & \ell_{3}\\
0 & 0 & 0
\end{array}
\right) b_{\ell_{1}\ell_{2}\ell_{3}}.
\]

The size of the VSA primary beam in the compact configuration
implies that the flat-sky approximation is not entirely accurate.
However, the correction factor is small, as described in 
\citet{Hobson}. The correction can be made be redefining the 
primary beam, and this is done in the power spectrum analysis.
Our current bispectrum calculation does 
not directly take this effect into account, but we anticipate the
associated error in the approximation to be small.


\section{The Very Small Array}
\label{sec:VSA}

The VSA is a 14-element interferometer situated on Mount
Teide in Tenerife. It has the ability to make observations
anywhere in the frequency range 26--36~GHz with an observing bandwidth 
of 1.5~GHz. During the first observing season (September 2000 --
September 2001) the antennae, of FWHM 4\fdg6, were arranged in
a compact configuration, with a maximum baseline of 1.5 m, 
and observations were made at 34~GHz.
In this configuration the VSA was sensitive to angular scales 
corresponding to $\ell\sim$ 150--900.
The VSA was then reconfigured to form the extended array, with
antennae of FWHM 2\fdg05 separated by a maximum distance
of 2.5 m, sensitive to the range $\ell\sim$ 300--1400. It has been
making observations in this configuration since October 2001,
at 33--34~GHz.

The compact array spent most of its time observing three separate
regions of sky (labelled VSA1, VSA2 and VSA3).
Within each region it made overlapping (mosaiced)
observations of two or three fields (denoted by either no suffix,
or the suffixes A, B or -OFF). Details of the fields are given in
\citet{Taylor}, and the power spectrum calculated from these
observations is presented in \citet{Scott}.

The extended array made mosaiced observations 
of smaller fields within the same three regions of the sky
(labelled with the suffixes E, F and G), details of which
can be found in \citet{Grainge}. 
Here, we have studied the compact and extended array data that were
used to compute the CMB power spectrum presented in \citet{Grainge}.
Since completing the analysis here there has been a new release of
results from further extended array observations \citep{Dickinson}.
Further details of the VSA observational technique can be found in
\citet{Watson}.

\subsection{Interferometer measurements}
\label{sec:measurements}

An interferometer samples the Fourier transform of the sky intensity
multiplied by the primary beam.
The fact that an interferometer measures directly in Fourier space
makes interferometric data particularly suited to analysis in Fourier space.
The visibility measured by the interferometer can be expressed as
\begin{equation}
V(\bmath{u})=
\int \Delta I(\bmath{\hat{x}})\;
A(\bmath{\hat{x}}) \:
e^{2\pi i\bmath{u \cdot \hat{x}}}\: \mathrm{d^2} \bmath{\hat{x}}
+ N(\bmath{u}),
\end{equation}
where $\Delta I(\bmath{\hat{x}})$ is the intensity fluctuation,
$A(\bmath{\hat{x}})$ the primary beam and $N(\bmath{u})$ the noise
on baseline $\bmath{u}$. To a good approximation the beamshape can
be modelled as Gaussian with $A(\bmath{x}) = \exp({-|\bmath{x}|^2/2\sigma^2})$
\citep{Hobson}, which is the shape that we
have assumed when simulating the observations.
In Appendix \ref{sec:appb} we describe how this leads to the approximate
expression for the variance of the visibilities
\begin{equation}
\left \langle V(\bmath{u})V^{\ast}(\bmath{u}) \right \rangle
\approx \pi \sigma^2 f^2 C(u) + \sigma^2_{\bmath u},
\label{eq:vis_var}
\end{equation}
where 
\begin{equation}
f = T_0 \frac{\partial B(\nu,T)}{\partial T} \Big\vert _{T=T_0}
\end{equation}
is the conversion to specific intensity and $\sigma^2_{\bmath u}$
is the variance of the noise. Similarly, for the three-point function
we obtain
\begin{equation}
\left \langle V(\bmath{u}_1)V(\bmath{u}_2)
              V(\bmath{u}_3) \right \rangle
\approx f^3 \frac{2}{3}\pi \sigma^2 B(u_1,u_2,u_3),
\label{eq:3point}
\end{equation}
for $\bmath{u}_1 + \bmath{u}_2 + \bmath{u}_3 = 0$. This approximation
has been used to convert the measured three-point function to
a bispectrum in units of $\mu\mathrm{K}^3$ for the purposes of
plotting. In Appendix \ref{sec:appc}
we derive an exact expression for a general case with a Gaussian
beam which it is necessary to use when trying to quantify a true
level of non-Gaussianity and considering a particular non-Gaussian
signal.

The sample time per visibility measurement by the VSA is typically
64 seconds. Each field was observed for $\sim$100--200 hours, 
resulting in a very large number of individual
visibility measurements. To reduce the data to a manageable level, the
measurements are binned into square cells on the $uv$ plane, using
a maximum-likelihood method as described in \citet{Hobson}. Adjacent
cells are still highly correlated as the cell width is chosen to be
less than the width of the aperture function (the Fourier 
transform of the primary beam) so as not to lose information.

\subsection{Source subtraction}

The VSA regions were carefully chosen in order to have relatively
low dust contamination (using the dust maps of \citealt*{Schlegel}),
low Galactic free-free and synchrotron
emission (as predicted by the 408-MHz all-sky survey of \citealt{Haslam})
and to avoid bright radio sources \citep{Taylor}. Choosing to
observe at the higher end of the VSA frequency range
reduces the signal from Galactic foregrounds \citep{Taylor}.
The compact array fields were chosen so as to contain
no sources brighter than 0.5 Jy. In order
to eliminate contamination from point sources, which would
otherwise contribute to excess power, they are
observed and directly subtracted from the data.
The source subtraction strategy used was first to
survey the fields with the Ryle telescope in Cambridge at 15 GHz 
prior to observation with the VSA. Then, simultaneous 
with VSA observations, a single-baseline interferometer,
with a dish separation of 9 m,
situated alongside the VSA, is used to monitor the
sources identified in the 15-GHz survey. 
For the compact array, it was calculated that all
sources brighter than 80 mJy needed to be subtracted.
Therefore the survey with the Ryle telescope sought 
to identify all sources above 20 mJy at 15 GHz. The
power of the CMB is more sensitive to sources at
higher values of $\ell$, and so for the extended array
the fields were chosen, using the previous survey, to
contain no sources brighter than 100 mJy \citep{Grainge}.
The sensitivity of the Ryle survey was increased
to 10 mJy, and all sources brighter than 20 mJy
subtracted from the measured visibilities. 


\section{Calculation of the bispectrum}

\subsection{Method of estimating the bispectrum}
\label{sec:calcbi}

Our starting point for estimation of the bispectrum of the VSA data
is equation (\ref{eq:flat_bi}).
The delta function tells us that we should be interested only in sets
of three vectors in the $uv$ plane that form a closed triangle. The lengths
of the sides of the triangle give us the values of $\ell_{1}$, $\ell_{2}$
and $\ell_{3}$. It follows that we can make an estimate of 
$B(\ell_{1},\ell_{2},\ell_{3})$ 
by finding from the data all possible triangles with these sides.

There are two additional factors to consider:
\vspace {-0.5\baselineskip}
\begin{enumerate}
\item The finite beamwidth means that the measured visibility on baseline
$\bmath{u}$ is a convolution with the aperture function (the Fourier 
transform of the primary beam),
as shown by equation (\ref{eq:v(u)}), so the visibilities
are `smeared' over a region of diameter a few wavelengths on the $uv$
plane. Therefore it is not desirable to form separate bispectrum estimates
using nearby visibility points. It is preferable to form individual 
estimates from `bands' of points on the $uv$ plane,
else adjacent estimates will be highly correlated. This also means that
we can slightly relax the requirement that the three vectors must form a
closed triangle, as long as the deviation from a closed triangle is
significantly less than the size of the aperture function.
\item The quantity that we are interested in is the ensemble-averaged
value of $B(\ell_{1},\ell_{2},\ell_{3})$, however, since we are 
observing only a small region of
our Universe, we want to average over as many triangles as is reasonable
in order to reduce the variance of the estimate.
\end{enumerate}
\vspace {-0.5\baselineskip}
In a general non-Gaussian scenario we would expect $B(\ell_{1},\ell_{2},\ell_{3})$
to vary reasonably slowly with $\ell$; if it fluctuates too rapidly then the
signal will be `washed out' both by the convolution and by combining
nearby points into one estimate. Predictions for the primordial
bispectrum (see Section \ref{sec:fnl}) show a similar number of 
peaks to the power spectrum (if we fix two values of $\ell$ and vary
the third), however, the bispectrum fluctuates between positive
and negative values.

Therefore, following the method used in \citet{Santos} for the frequentist
estimator, we divide the $uv$ plane into concentric annuli, of width
$\Delta$. To form a bispectrum estimate, three annuli are selected, and we
set $\ell_{1}$, $\ell_{2}$ and $\ell_{3}$ equal to the radii of the midpoints
of the annuli. The data are binned on a square grid in the $uv$ plane. This
binning shifts the points very slightly and hence is equivalent to relaxing
the requirement of exact triangles.
We find all the possible triangles formed from vectors 
$\bmath{u}_{1}$,
$\bmath{u}_{2}$ and $\bmath{u}_{3}$ where
$\bmath{u}_{i}$ points 
to the centre of a cell containing a data point in annulus $i$.
 We can express our estimator as:
\begin{equation}
\hat{B}(\ell_{1},\ell_{2},\ell_{3})
= \frac{\sum _{\bmath{u}_{i}\in A_{i}}
      \frac{V(\bmath{u}_{1})V(\bmath{u}_{2})V(\bmath{u}_{3})}
	   {\sigma ^{2}_{\bmath{u}_{1}\bmath{u}_{2}\bmath{u}_{3}}}}
      {F \sum \frac{1}{\sigma ^{2}_{\bmath{u}_{1}\bmath{u}_{2}\bmath{u}_{3}}}}.
\label{eq:estimator}
\end{equation}
where $F = f^3 \frac{2}{3}\pi\sigma^2$, the prefactor in equation
(\ref{eq:3point}), which accounts for the conversion to flux density
and the effect of the primary beam.
The $\sigma ^{2}$ are used to weight each individual triangle of vectors,
and are estimated as explained in Section \ref{sec:weight}. 
$A_{i}$ denotes annulus $i$.
The visibilities are complex, but the resulting bispectrum estimates
are real since $a(-\bmath{u})=a^{*}(\bmath{u})$,
by virtue of the fact that the temperature field is purely
real.\footnote{Although the signal contribution to the visibilities
satisfies $a^*(\bmath{u})=a(-\bmath{u})$, the noise is generally
uncorrelated between visibilities. In our analysis we impose
the symmetry $a^*(\bmath{u})=a(-\bmath{u})$ on the noise too
by averaging the data $a(\bmath{u})$ and $a^*(-\bmath{u})$ to form the
visibility at $\bmath{u}$.}

Our method for estimating the bispectrum from interferometer data
mirrors that used for the {\sl MAXIMA} data \citep{Santos}.
However, the {\sl MAXIMA} data are obtained in real space and so it is necessary
first to perform a Fourier transform. We begin with the data in the
$uv$ plane, so we do not have to concern ourselves with window functions.
However, we have to deal with non-uniform $uv$ plane coverage, which
results in large variations in the noise on each cell.
There is also a lot of variation in the number of triangles of vectors
which form each bispectrum estimate. 

\subsubsection{Choice of $\Delta$}

\begin{figure}
\centering
\includegraphics[width = \linewidth]{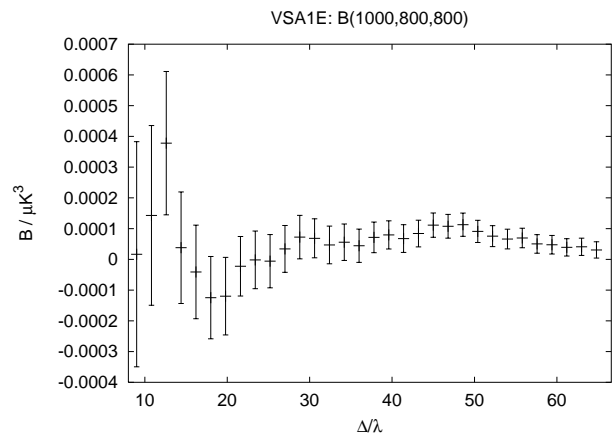}
\caption{\label{fig:delta}Illustration of the variation of a bispectrum
estimate with \protect\( \Delta \protect \), the width of the annuli,
for the extended array 
field VSA1E, with estimated variance calculated according to equation
(\ref{eq:var_est}).
For small values, there are few triangles and so the estimate is large,
and fluctuates rapidly. The bispectrum appears to converge towards
zero as we increase \protect\( \Delta \protect \), and varies smoothly
with \protect\( \Delta \protect \).
We wish to chose the smallest value of \protect\( \Delta \protect \) 
for which the bispectrum estimate is changing reasonably slowly,
and for which adjacent bispectrum estimates are uncorrelated.}
\end{figure}

A natural scale for $\Delta$, the width of the annuli, is set by the
width of the aperture
function. Increasing $\Delta$ increases the number of triangles
that form each bispectrum estimate, but they correspond to a wider
spread of the underlying values of $\ell$. This loss of $\Delta \ell$
resolution runs the risk of `washing out' any oscillatory signal
that may be present. However, if $\Delta$ is too small then the estimates
have large variances resulting from there being few
triangles forming each estimate. In addition, adjacent estimates
would be highly correlated. Therefore we chose a 
value of $\Delta$ which is large enough to encompass the width of the
aperture function, but with the relative size compared to the aperture
function slightly smaller for the case of the extended array, since
it has a larger aperture function. Fig.~\ref{fig:delta} shows 
how particular bispectrum estimates vary with the
width of the annuli. For the compact array we chose $\Delta$ = 16 $\lambda$,
and for the extended array, $\Delta$ = 32.4 $\lambda$.
This is equivalent to 1.46 and 1.31 times the FWHM of the aperture function
for the compact and extended arrays respectively.
\subsubsection{Weighting the bispectrum estimates}
\label{sec:weight}

We can achieve a near-optimal estimator by weighting each visibility 
with its inverse variance. In the absence of primary beam effects,
this estimator would be the optimal cubic estimator of the bispectrum
(see Appendix \ref{sec:appa}).
We thus weight each individual triangle as 
$\frac{1}{\sigma _{\bmath{u}_{1}\bmath{u}_{2}\bmath{u}_{3}}^{2}}$
where  
\begin{equation}
\sigma _{\bmath{u}_{1}\bmath{u}_{2}\bmath{u}_{3}} ^{2}
=\left( \widetilde{C}_{\ell_1} + \sigma _{\bmath{u}_{1}}^2 \right)
 \left( \widetilde{C}_{\ell_2} + \sigma _{\bmath{u}_{2}}^2 \right)
 \left( \widetilde{C}_{\ell_3} + \sigma _{\bmath{u}_{3}}^2 \right),
\label{eq:weight}
\end{equation}
is the product of the variances of the visibilities. Here,
$\widetilde{C}_{\ell_i}$ is given by $\pi\sigma^2f^2C(\bmath{u}_i)$ as
shown in equation (\ref{eq:vis_var}). 

We can make a rough estimate of the variance of the bispectrum
by ignoring correlations due to the primary beam, and neglecting
the fact that some visibilities are used more than once in the
bispectrum estimate. Under the assumption
of a Gaussian signal, each triangle is the product of three
independent visibilities, and nearly all triangles in a given
estimate $\hat{B}$ are also independent. It follows that
\begin{equation}
\sigma _{\hat{B}}^{2}
\approx \frac {1} {F^2 \sum_{\bmath{u}_i \in A_i} \frac{1}
	{\sigma^{2}_{\bmath{u}_1 \bmath{u}_2 \bmath{u}_3}}} .
\label{eq:var_est}
\end{equation}
In practice, we compute errors and assess the statistical significance
of our results using Gaussian simulations which properly take account
of correlations, however, we can use equation (\ref{eq:var_est}) as a
way of checking the calculation of the weights as given by equation
(\ref{eq:weight}).

\subsection{Simulations}
\label{sec:sims}

\subsubsection{Gaussian Simulations}

\begin{figure*}
\centering
\begin{minipage}{150mm}
\centering
\includegraphics[width=0.32\textwidth]{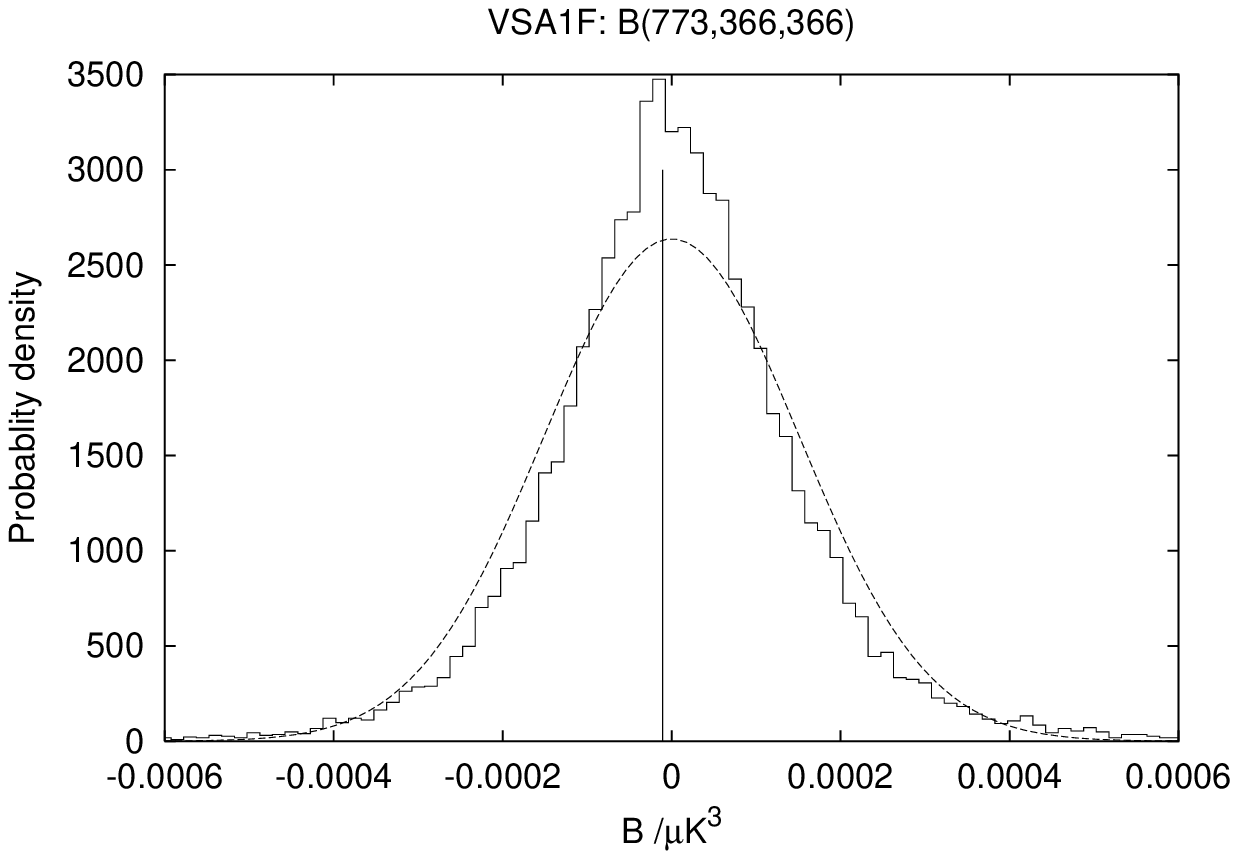}
\includegraphics[width=0.32\textwidth]{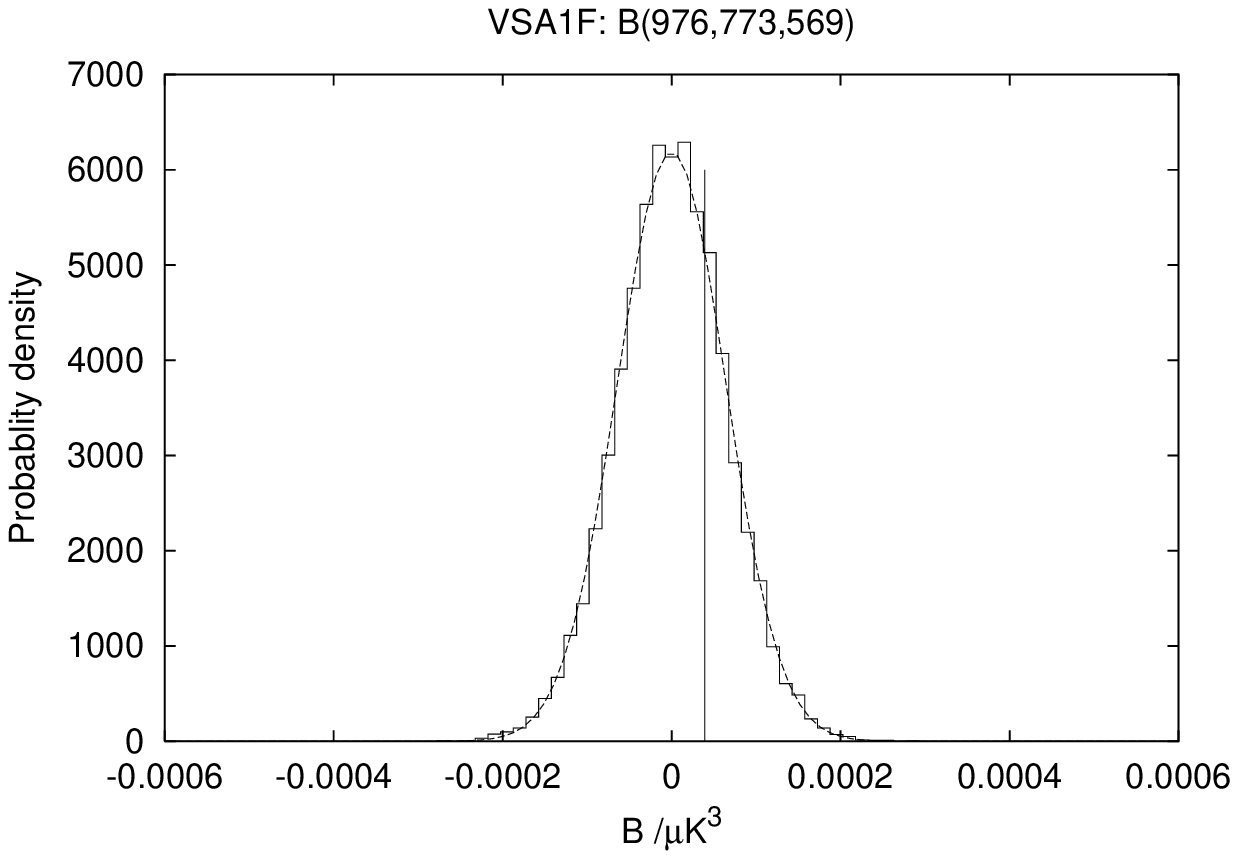}
\includegraphics[width=0.32\textwidth]{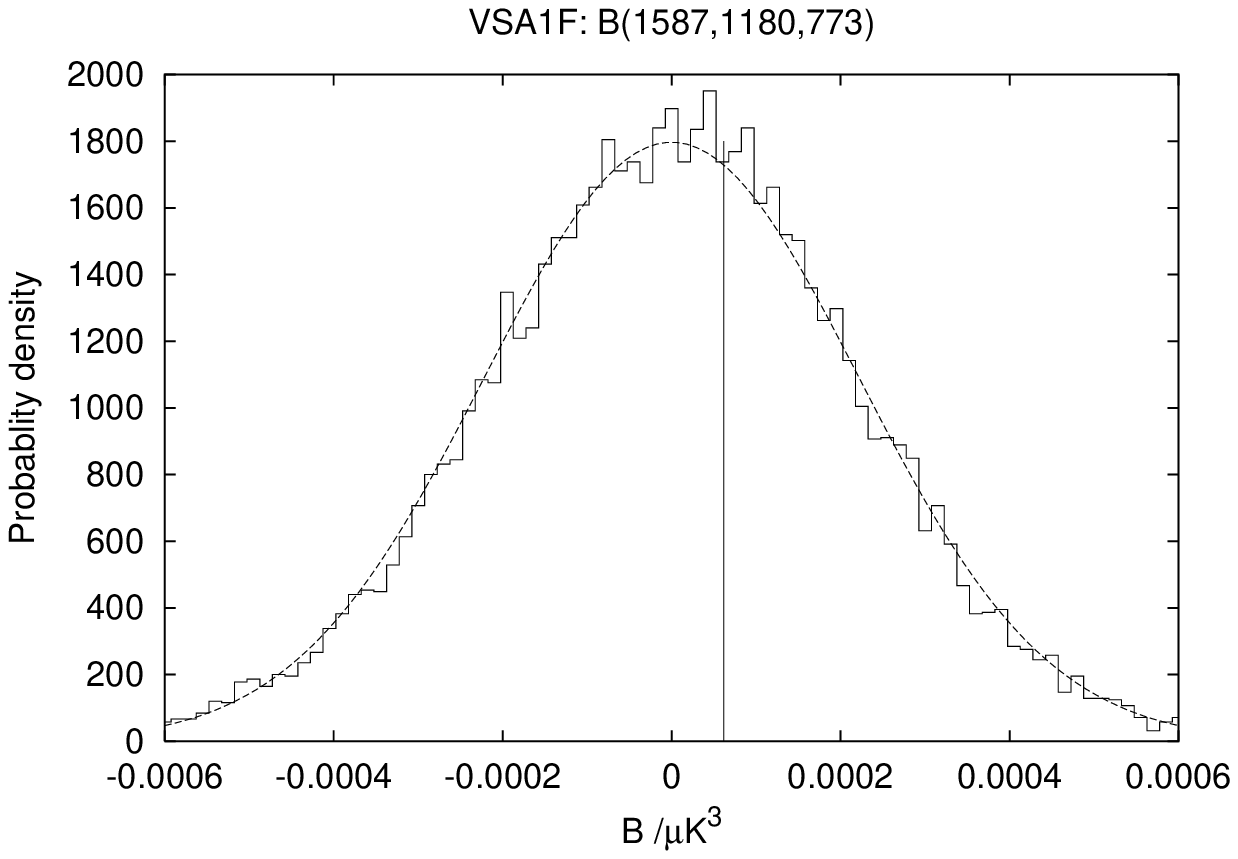}
\caption{\label{fig:dist}Distribution of bispectrum estimates for VSA1F
obtained after 15000 simulations, and comparison with a Gaussian
distribution with the same variance. The measured value from the real
data is shown by a vertical line.}
\end{minipage}
\end{figure*}

In order to assess the statistical significance of our results, we 
have performed Gaussian simulations of the VSA data from a
simulated Gaussian sky. We extract visibilities at the same $uv$
positions as our data, and add noise which is drawn from a Gaussian
distribution with the same variance as the noise on the data. For each
simulation we compute the bispectrum, with the same code as that
employed on the real data. From the suite of 15000 simulations we estimate
the variance in each bispectrum estimate, and their covariances
(which arises from a combination of sample variance
and instrumental noise). 

The power spectrum for the simulations is drawn from a flat $\Lambda$CDM model,
with the same parameters as model A in \citet{Slozar}, using all CMB
data. Model A was the simplest and had the highest evidence.

Fig.~\ref{fig:dist} shows the distribution of some bispectrum estimates
obtained from the Gaussian simulations, together with the value calculated
from the real data. The estimates which are formed from
more triangles tend to have a smaller variance as would be expected. The
majority of the distributions fit well with a Gaussian curve with the
same variance, although there are exceptions (which tend to be when some of
the $\ell$s are low) such as the distribution of $B(773,366,366)$. 

\subsubsection{Non-Gaussian Simulations}
\label{sec:ng_sims}
It is desirable to test the bispectrum estimator on simulated data
with a known bispectrum to check for bias. In order to do this, 
we adopt a probability distribution function (pdf) derived from the Hilbert
space of a linear harmonic oscillator, developed by \citet{rocha}.
This exercise helps us to assess the performance of our code to compute
the bispectrum, and could also be a useful alternative
hypothesis in testing for non-Gaussianity.
We give here a brief account of the procedure followed to simulate these
non-Gaussian CMB maps; a more detailed description will be presented in
Rocha et~al. (in preparation).

We start by drawing the values of the CMB temperature fluctuations, 
$\Delta T(\bmath{\hat x}_p)$, independently in each real-space
pixel from our non-Gaussian pdf. Since
the pdf is based on the wavefunctions of the eigenstates of a linear harmonic
oscillator, it takes the form of a Gaussian multiplied by the square of a
(possibly finite) series of Hermite polynomials where the coefficients 
$\alpha_{n}$ are used as non-Gaussian qualifiers.
These amplitudes $\alpha_{n}$ can be written as series of cumulants 
\citep*{joao}, and can be 
independently set to zero without mathematical inconsistency \citep{rocha}.
Hence we should regard $\alpha_n$
as non-perturbative generalisations of cumulants. 
Let $x$ represent a general random variable, within a set of variables
which are assumed to be independent. 
The most general probability density for the fluctuations  
in $x$ is thus:
\begin{equation} 
P(x)=|\psi(x)|^2= e^{-{x^2\over 2 \sigma_0^2}}
\left|\sum_n
\alpha_n C_n H_n{\left(x\over {\sqrt 2}\sigma_0\right)}\right|^2,
\end{equation}
where the $H_n$ are the Hermite polynomials, and the quantity
$\sigma_0^2$ is the variance associated with the (Gaussian) probability 
distribution for the ground state $|\psi_0|^2$. The $C_n$ are fixed
by normalising the individual states.
The only constraint on the amplitudes $\alpha_n$ is:
\begin{equation}\label{eq:const}     
\sum |\alpha_n|^2 = 1.
\end{equation}
This is a simple algebraic expression which can be eliminated explicitly
by setting $\alpha_0= \sqrt{1- \sum_1^\infty |\alpha_n|^2 }$.

\begin{figure}

\begin{center}
\psfig{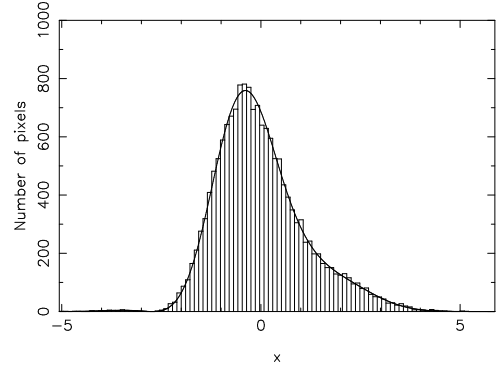}
\end{center}

\caption{Non-Gaussian pdf given by equation (\ref{eq:like}) with 
$\alpha_{1}=\alpha_{2}=0$, $\alpha_{3}=0.2$ and $\sigma_{0}=1$.}
\label{fig:sho} 

\end{figure}

We consider here the situation in which all $\alpha_n$ are set to zero,
except for the real part of $\alpha_3$ (and consequently $\alpha_{0}$).
The reason for this is that such a quantity reduces to the skewness in 
the perturbative regime. 
The imaginary part of $\alpha_3$ is only meaningful in the non-perturbative
regime (and can be set to zero independently without inconsistency;
\citealt{rocha}). Hence we are considering a pdf of the form:
\begin{equation} \label{eq:like}
P(x)= \frac{e^{-x^{2}/(2 \sigma_{0}^{2})}}{\sqrt{2 \pi}
\sigma_{0}} 
\left[ \alpha_{0} + \frac{\alpha_{3}}{\sqrt{48}} 
H_{3} \left(\frac{x}{\sqrt{2} \sigma_{0}}\right) \right]^{2}, 
\end{equation}
with $\alpha_0= \sqrt{1- \alpha_3^{2}}$.
This allows us to obtain a centred distribution with $\mu_{1}=0$ , 
where $\mu_{n}$ is the $n$th moment around the origin defined as:
\begin{equation}
  \mu_n = \langle x^n \rangle = \int^{\infty}_{-\infty} x^nP(x)\mathrm{d}x.
\end{equation}
The first, second and third moments of our pdf are related to $\alpha_3$ 
and $\sigma_{0}$ as follows \citep{Contaldi}:
\begin{eqnarray}
\mu_1&=&0,  \nonumber\\
\mu_2&=&\sigma_0^{2}\left( 1+6\alpha_3^{2} \right), \nonumber\\
\mu_3&=&\left( 2\sigma_0^{2} \right)^\frac{3}{2} 
	\sqrt{3 \left[ \alpha_3^{2} \left(1-\alpha_3^{2} \right) \right]}.
\end{eqnarray}
Therefore we have generated a centred distribution with a fixed variance and
skewness.
For our purposes here we considered the distribution with $\alpha_{3}=0.2$
and $\sigma_{0}=1$ plotted in Fig.~\ref{fig:sho}.

The space of possible distribution functions is constrained due
to restricting the set of parameters to two parameters only.
This implies that we cannot generate distributions with any
given variance and skewness. However, in general our method can
generate higher values of the relative skewness (since it can
generate any distribution) but for that purpose one needs
more parameters $\alpha_n$ \citep{Contaldi}.

\begin{figure}
\begin{center}
\includegraphics[height= 0.2\textheight,angle = 270]{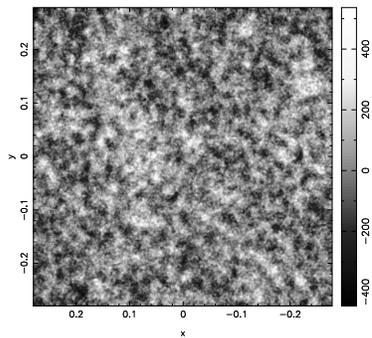}
\end{center}
\caption{A Non-Gaussian simulation of the sky in real space obtained 
using the procedure described in Sec. \ref{sec:ng_sims}}
\label{fig:ngsim}
\end{figure}

The maps simulated by drawing the pixel values from this pdf
will be, by construction, statistically isotropic.
They consist of non-Gaussian white noise
with variance given by $\mu_2$. We then Fourier transform the map
$\Delta T(\bmath{\hat x}_p)$ to get the Fourier modes, $a(\bmath u)$,
which have variance proportional to $\mu_2$. We rescale
these Fourier coefficients so that the variance is given by the correct
angular power spectrum $C_{\ell}$.
We then inverse Fourier transform these coefficients back to real space to
produce a new signal map, $\Delta T(\bmath{\hat x}_p)$, which now 
has the appropriate covariance matrix. In Fig.~\ref{fig:ngsim} 
we plot one of these non-Gaussian maps. This new map can then be used
as input to simulate the VSA observational strategy and to obtain
a set of visibilities as observed by the VSA.
For our purposes we are interested in the simplest case, i.e.\ with no beam
convolution and no noise, though these can easily be incorporated if we 
wish. We output the visibilities on a square grid, and then compute the
bispectrum of these simulated visibilities with the same code that
we apply to the real data. 

\begin{figure}
\begin{center}
\includegraphics[height = 0.2\textheight]{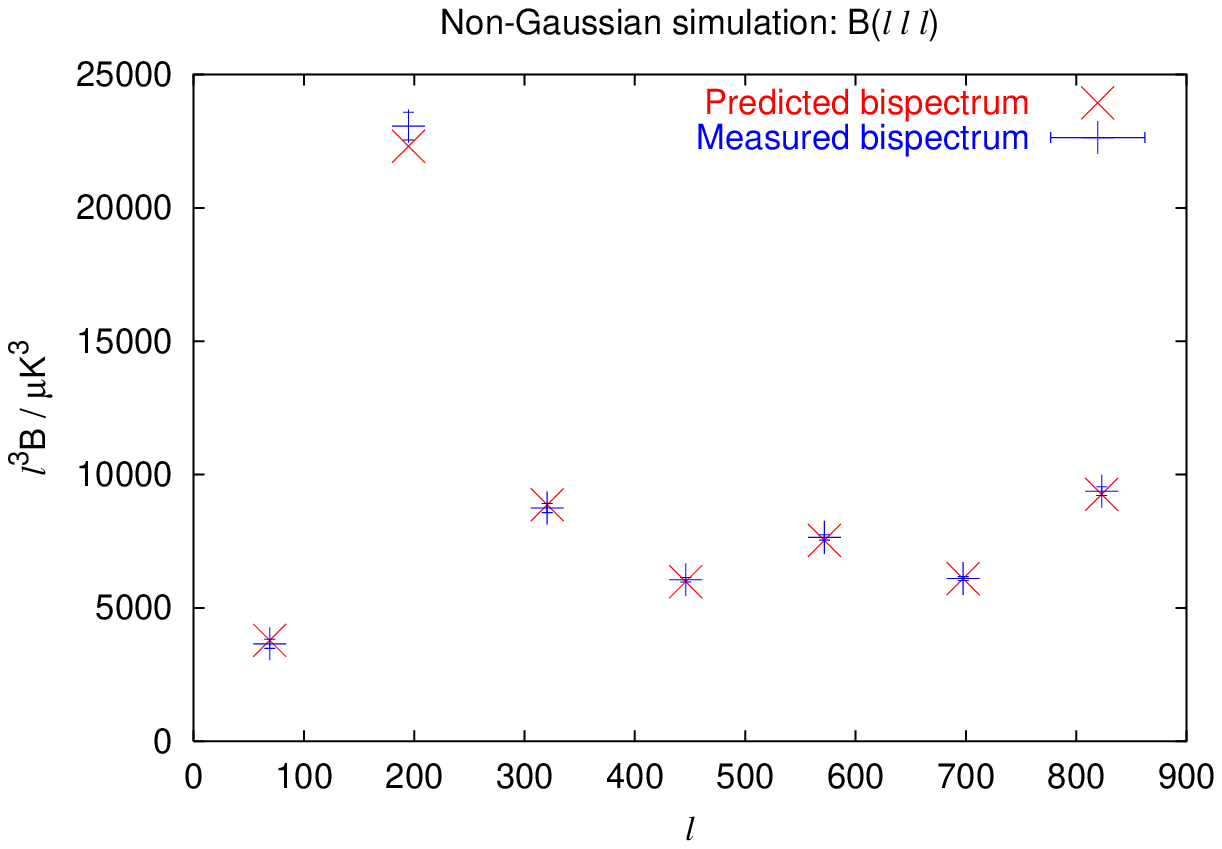}
\includegraphics[height= 0.2\textheight]{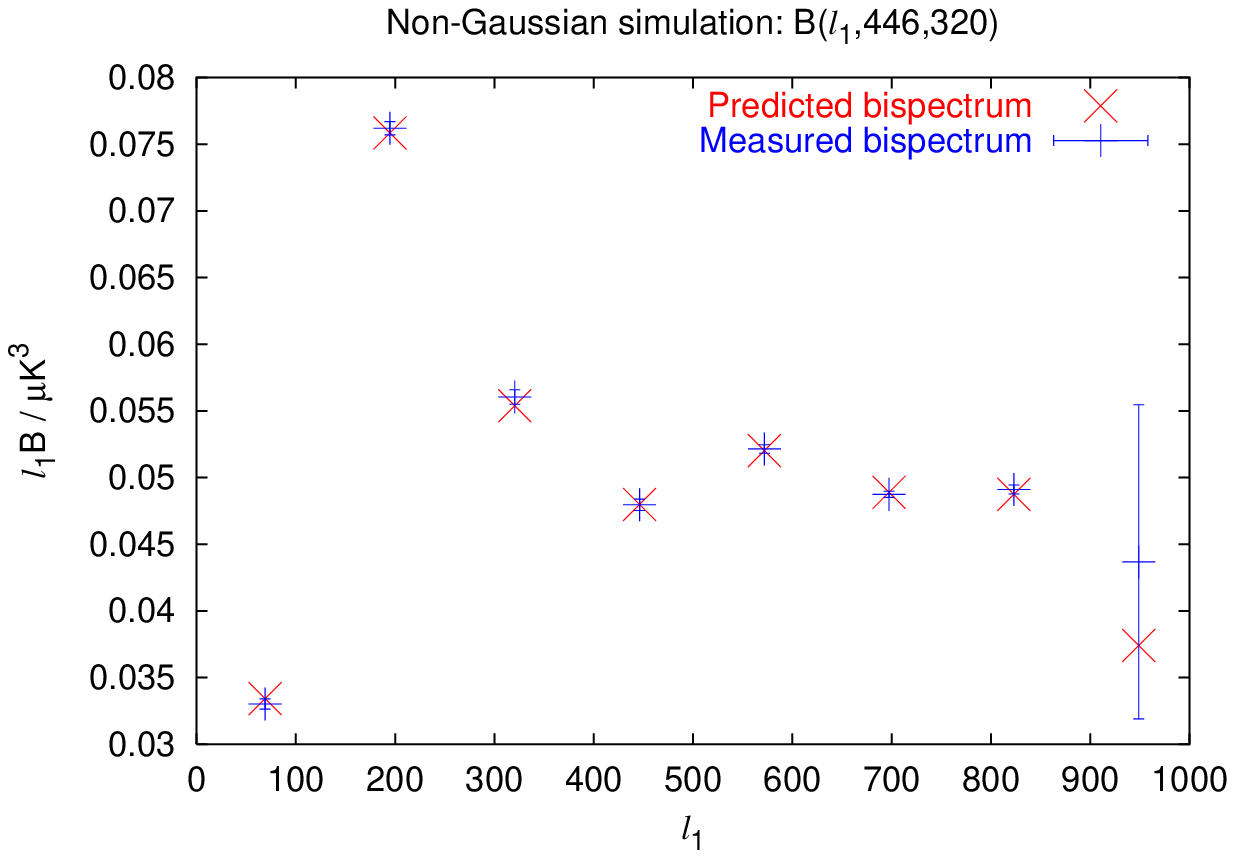}
\end{center}

\caption{Bispectrum estimates of the non-Gaussian simulations
with $\alpha_{1}=\alpha_{2}=0$,  $\alpha_{3}=0.2$ and $\sigma_{0}=1$
(i.e. with $\mu_{3} \sim 0.95$ for a patch of the sky of side 28\fdg6
and pixel number $N_{\mathrm{pix}}= 128^2$), as compared to the predicted 
bispectrum obtained using equation (\ref{eq:ngbi}). The error on the 
measured bispectrum are due to the finite number of simulations (1865).}
\label{fig:biskew}
\end{figure}

By construction, the power spectrum of the non-Gaussian map is $C_{\ell}$
and the bispectrum is related to the 
skewness and variance of our non-Gaussian pdf by:
\begin{equation}
b_{\ell_1 \ell_2 \ell_3}= 
\frac{\mu_{3}}{\mu_{2}^{3/2}} A^{1/2}_{\mathrm{pix}} 
\sqrt{ C_{\ell_1} C_{\ell_2}C_{\ell_3}}
\label{eq:ngbi}
\end{equation}
where $A_{\mathrm{pix}}$ is the pixel area given by 
$A_{\mathrm{pix}}= L^2 / N_{\mathrm{pix}}$,
for a small patch of the sky of area $L^{2}$.
For a detailed calculation see Rocha et~al.~(in preparation).
In Fig.~\ref{fig:biskew} we plot the computed values of the bispectrum against
the predicted ones. The agreement shows that our bispectrum calculations are 
indeed correctly obtained.

We note that from the generalised Bayesian analysis using the non-Gaussian 
likelihood developed by \citet{rocha}, and applied to the VSA 
data by \citet{Savage}, we concluded that VSA data are mostly 
consistent with zero $\alpha_{3}$, resulting in no evidence for this
type of non-Gaussianity.

\subsection{Testing for non-Gaussianity}
\label{sec:tests}

\begin{figure}
\centering 
\includegraphics[width=\linewidth]{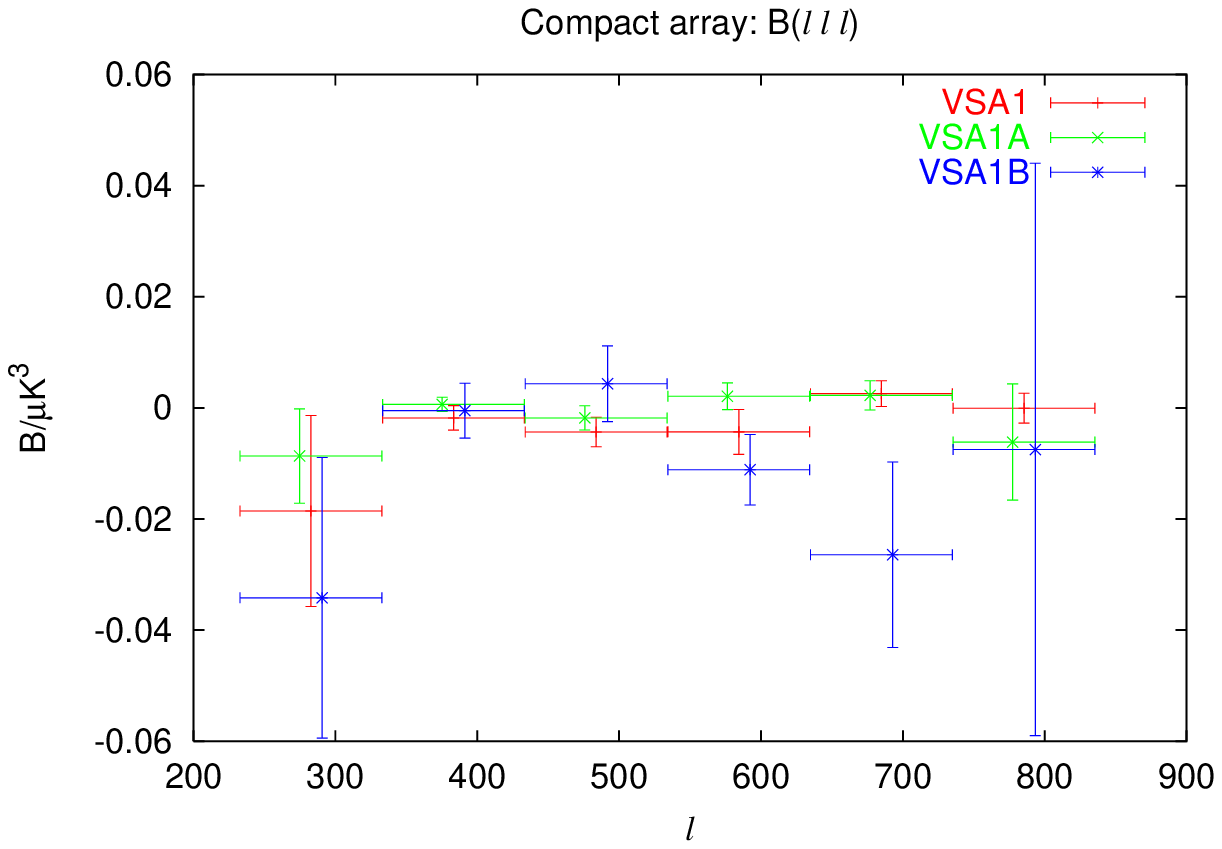}
\includegraphics[width=\linewidth]{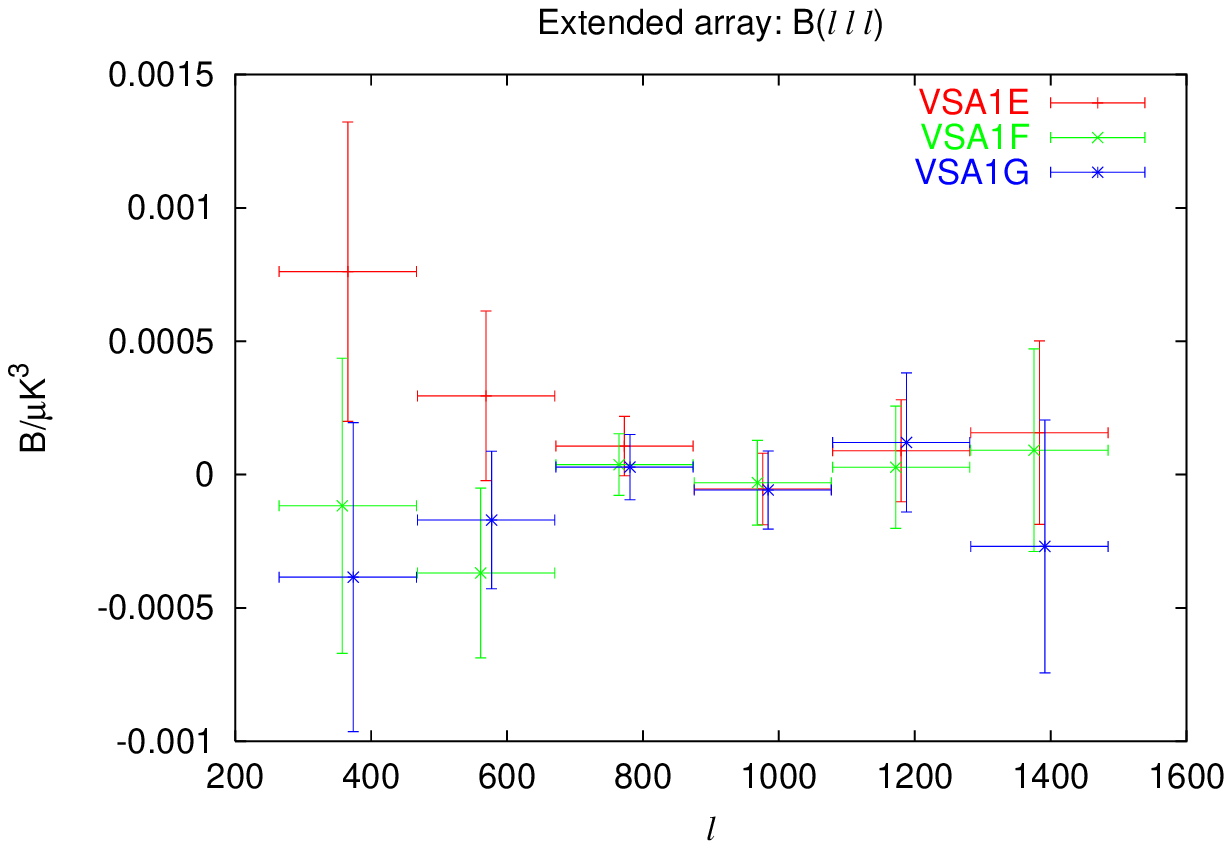}
\caption{\label{fig:bispec}Estimated diagonal component of 
bispectrum from the compact (top) and extended (bottom)
arrays for the region VSA1, with error bars from Gaussian
simulations.
Some values with very large errors at high and low $\ell$ have
been omitted.}
\end{figure}

Fig.~\ref{fig:bispec} shows the estimates of $B(\ell,\ell,\ell)$, together
with error estimates from Gaussian simulations, from the observations of the
three compact array fields and the three extended array fields
in the VSA1 region. Little correlation 
can be seen between the three datasets in each graph.
At low values of $\ell$ noisy estimates are obtained due to the
fact that few triangles can be found at low $\ell$. The estimated
variance of the mean is not dissimilar to the mean itself, indicating
that we cannot expect to find a significant deviation from zero. The
bispectrum estimates from the extended array are much less noisy than 
those from the compact array, reflecting the fact that the noise on the 
visibilities is smaller for the extended array than for the
compact array (the mean noise per binned visibility is 4.4--8.8 Jy
for the compact array and 0.7--1.1 Jy for the extended array).
The noise contributes as $\sigma_{\bmath{u}}^3$ to the bispectrum error.

\begin{figure}
\centering 
\includegraphics[height = 0.2\textheight]{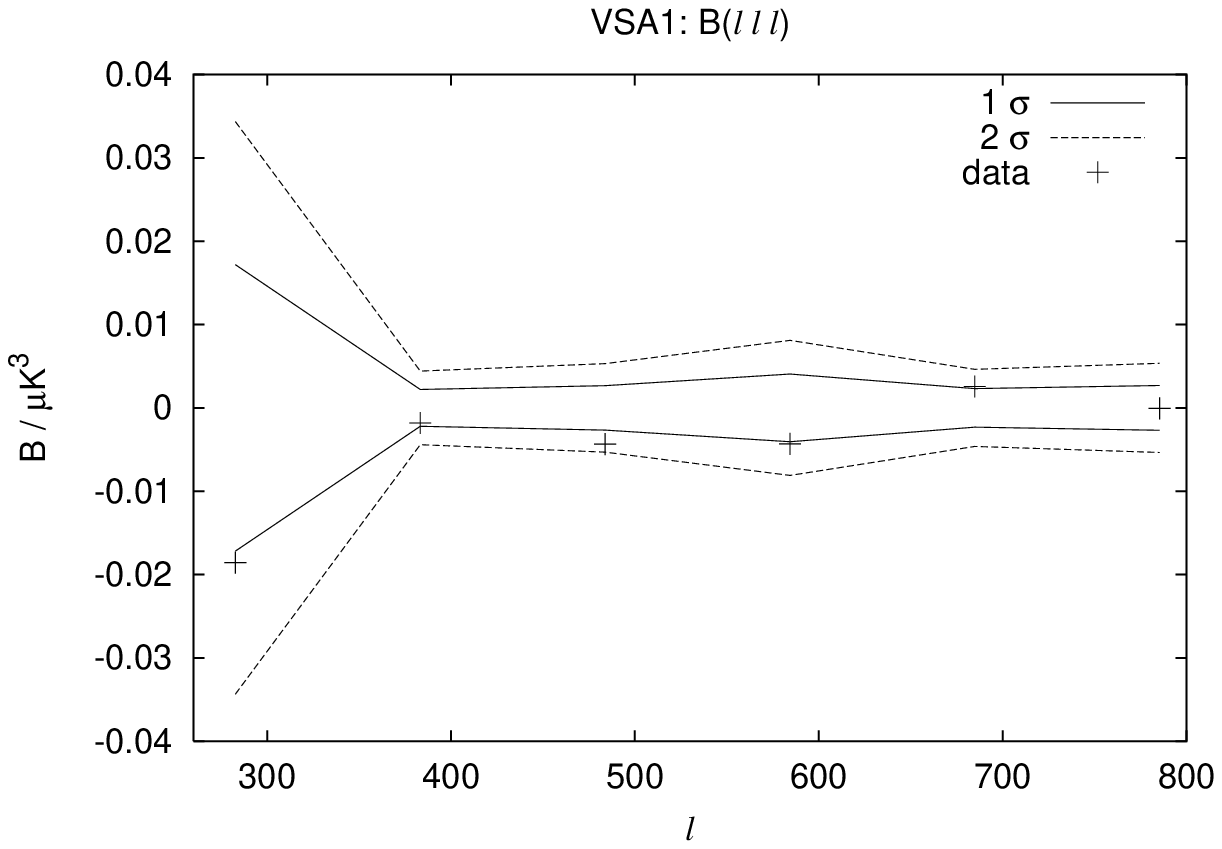}
\includegraphics[height = 0.2\textheight]{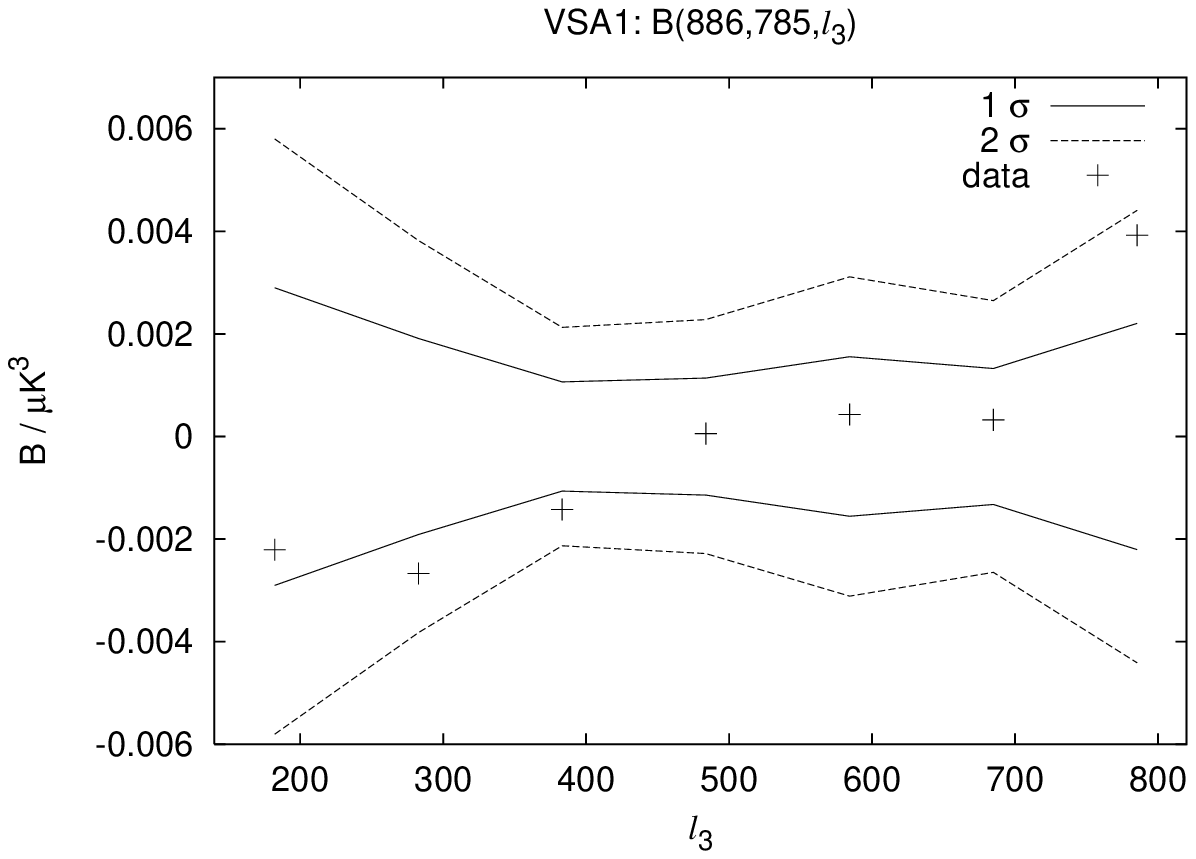}
\caption{\label{fig:1A}Results from simulations for VSA1 showing
variance of bispectrum estimates and data values. The top figure
shows the diagonal component of the bispectrum.}
\end{figure}

\begin{figure*}
\centering 
\begin{minipage}{150mm}
\includegraphics[width = 0.45\textwidth]{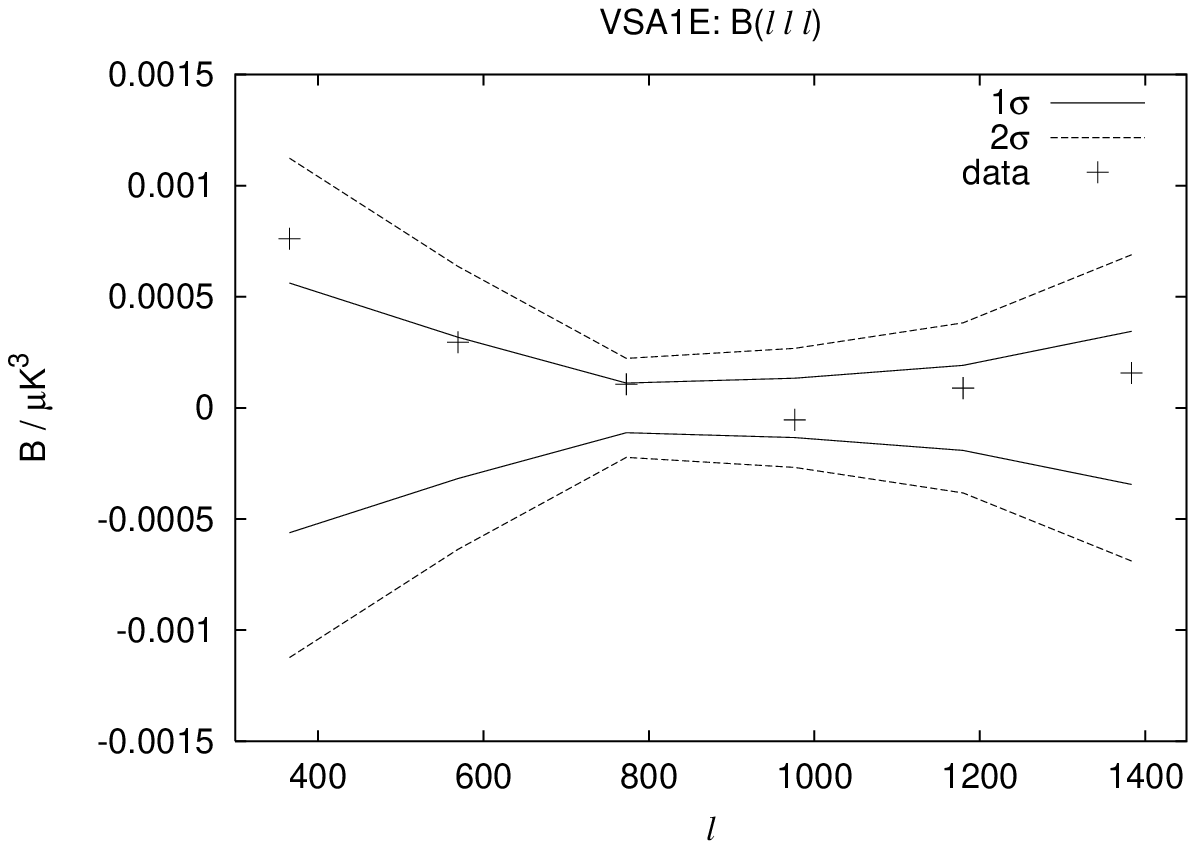}
\includegraphics[width = 0.45\textwidth]{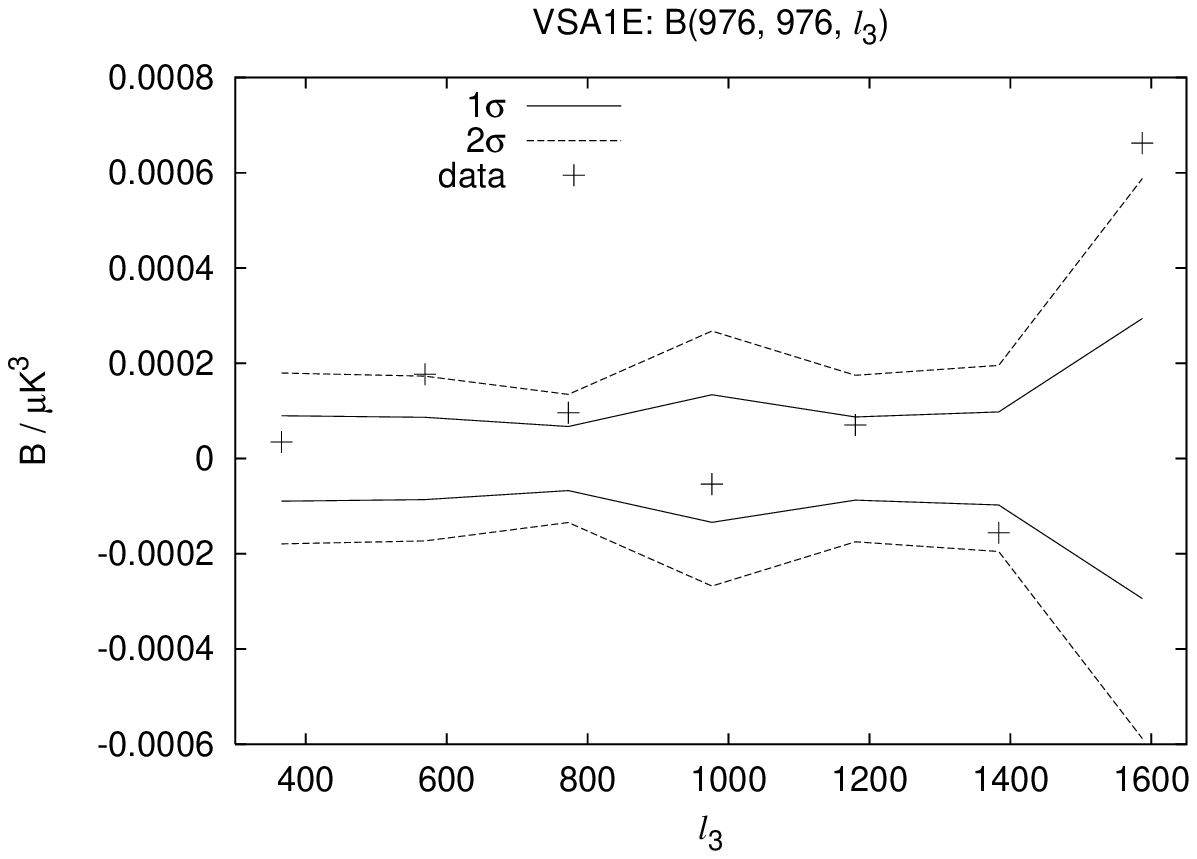}
\end{minipage}
\begin{minipage}{150mm}
\includegraphics[width = 0.45\textwidth]{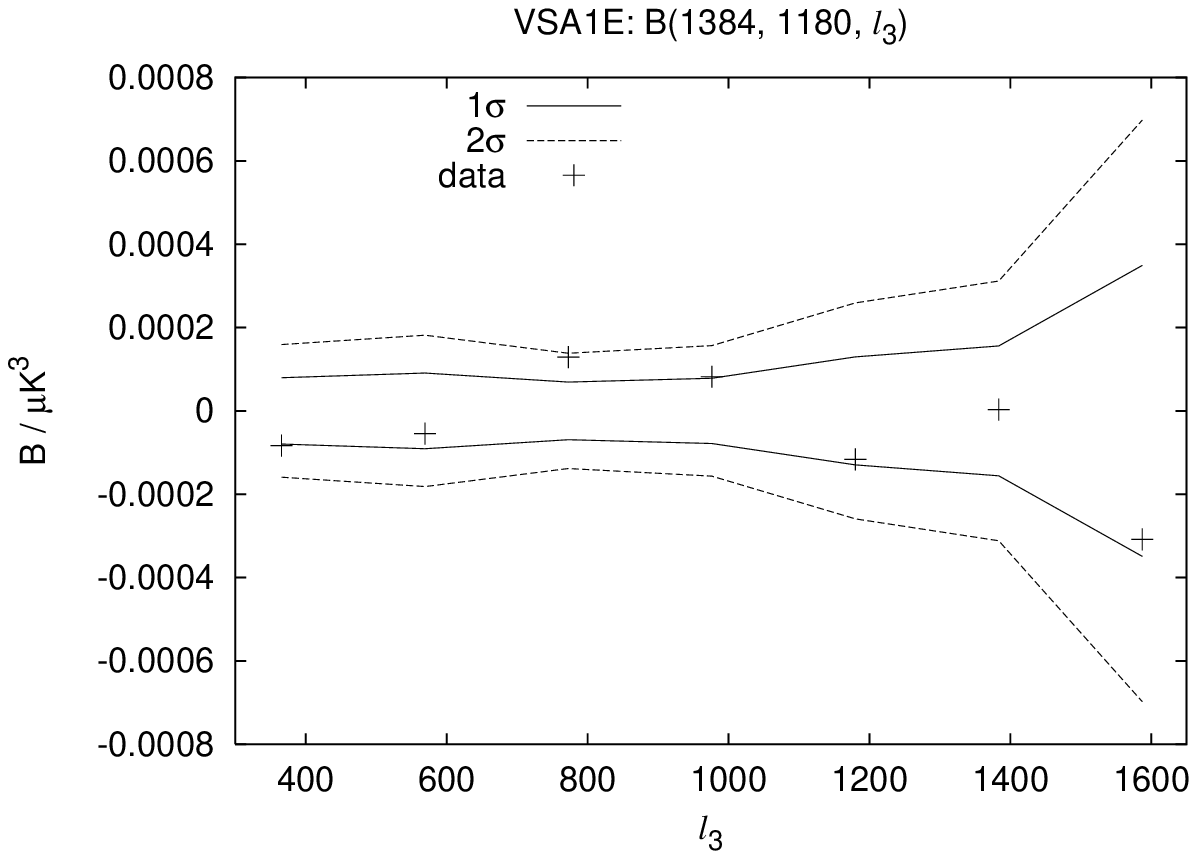}
\includegraphics[width = 0.45\textwidth]{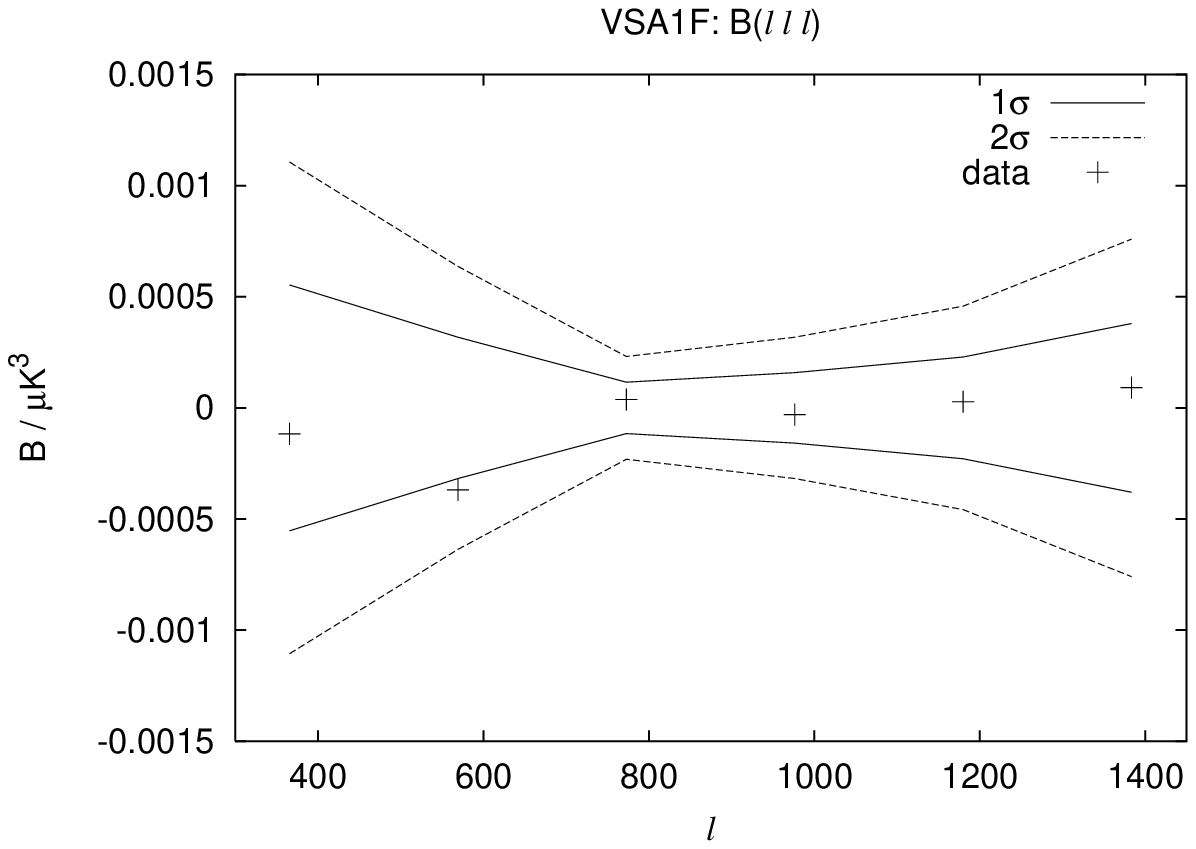}
\end{minipage}
\caption{\label{fig:1E}Bispectrum estimates with variance from simulations
for the extended array fields VSA1E and VSA1F.}
\end{figure*}

\begin{figure}
\centering
\includegraphics[height = 0.2\textheight]{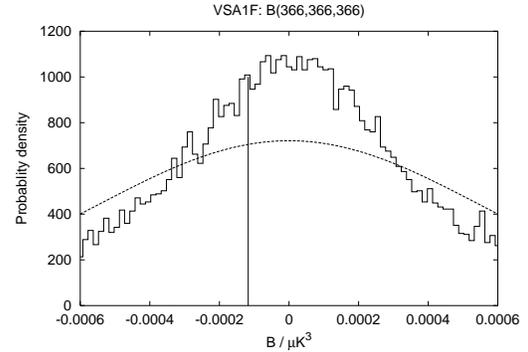}
\caption{\label{fig:B366} Distribution of the bispectrum estimate
$B(366,366,366)$ for the field VSA1F, and comparison with a Gaussian
distribution with the same variance.}
\end{figure}

Figs~\ref{fig:1A} and \ref{fig:1E} show a sample of the
bispectrum values computed from
the real data, together with the 1 and 2-$\sigma$ values from the
Gaussian simulations. The change in the variance with multipole is 
due mainly to the variation in the number of triangles of vectors
used to form each bispectrum estimate. In addition, the noise on
each binned visibility value varies as a result of
the differing numbers of raw visibility measurements used to form
each binned value. 
The rough estimates of the errors from equation (\ref{eq:var_est})
agree well with the 1-$\sigma$ values from the simulations for large
values of $\ell$, tending to be too small by approximately 0--4 per cent.
However, for small values of $\ell$, equation
(\ref{eq:var_est}) generally underestimates the variance by a greater 
amount, the most extreme example being for $B(366,366,366)$
where the estimated variance is only 35 per cent of the value 
obtained from simulations for the case of the VSA1F field. The
distribution of this particular bispectrum value is very non-Gaussian, as
shown in Fig.~\ref{fig:B366}, and hence it appears that 
the correlations that we have neglected in equation (\ref{eq:var_est})
are significant in this case.
Although around 160 different triangles of vectors in the $uv$ plane
make up the bispectrum estimate, the effective number of independent
triangles is much smaller.

For the 17 fields, there are a total of 1839 estimated bispectrum
values. Of these, 99 were found to be greater than 2$\sigma$ in
magnitude, compared with the expected value of $\sim$84 if the
bispectrum estimates were Gaussian distributed (which we have
found to be the case except for when one of the $\ell$s is small).
The VSA3 field had a significant number of large bispectrum 
estimates (18 out of 176 of modulus $>2\sigma$, 6 of which were
$>3\sigma$), most of the largest of which appear to have in
common at least one value of $\ell$=584. There was one bispectrum
estimate outside the 4-$\sigma$ limit, in the field VSA2G.
In order to proceed further, we need a way of testing the
data as a whole. In the following section we describe one such
test that we have applied to the individual fields, as well as
to sets of bispectrum estimates formed from the weighted sum
of the estimates for individual fields that are in the same region
of the sky.

\subsubsection{Kolmogorov-Smirnoff test}

\begin{figure}
{\centering 
\includegraphics[height=0.2\textheight]{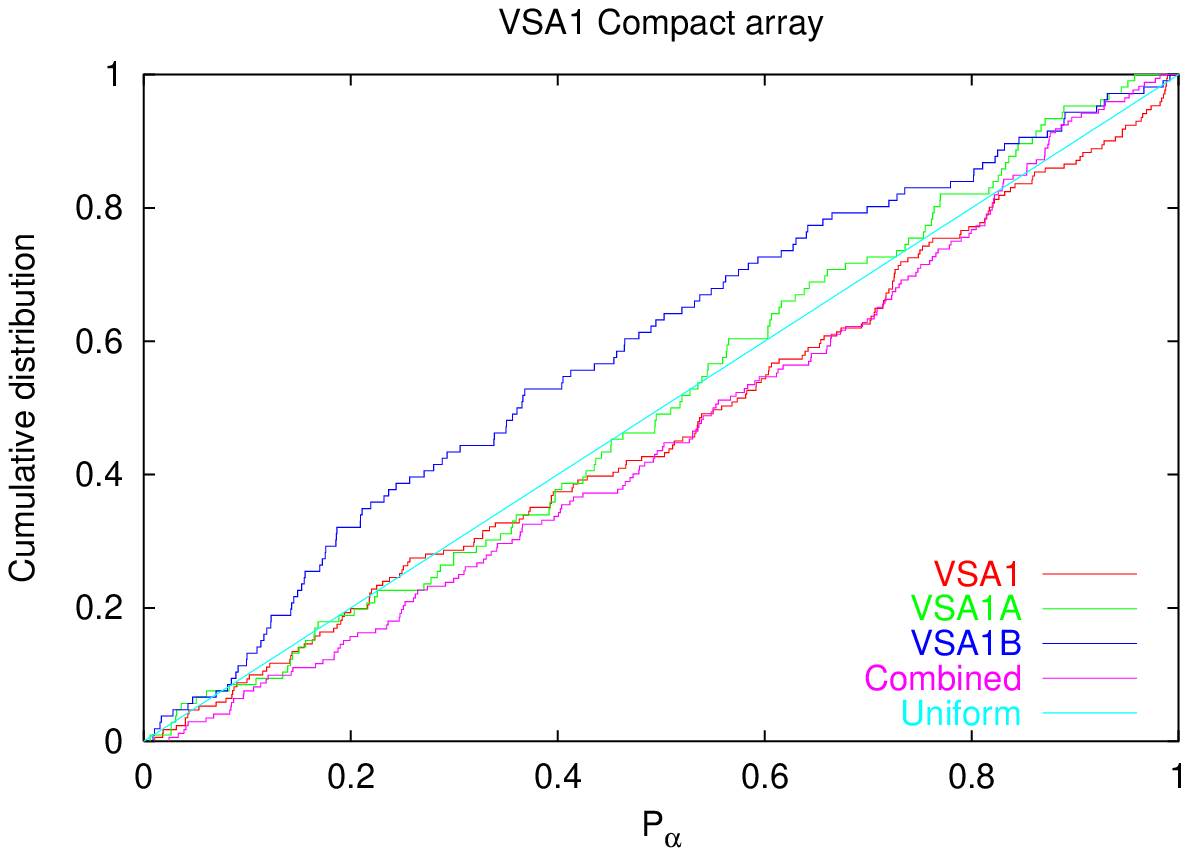}  
\includegraphics[height=0.2\textheight]{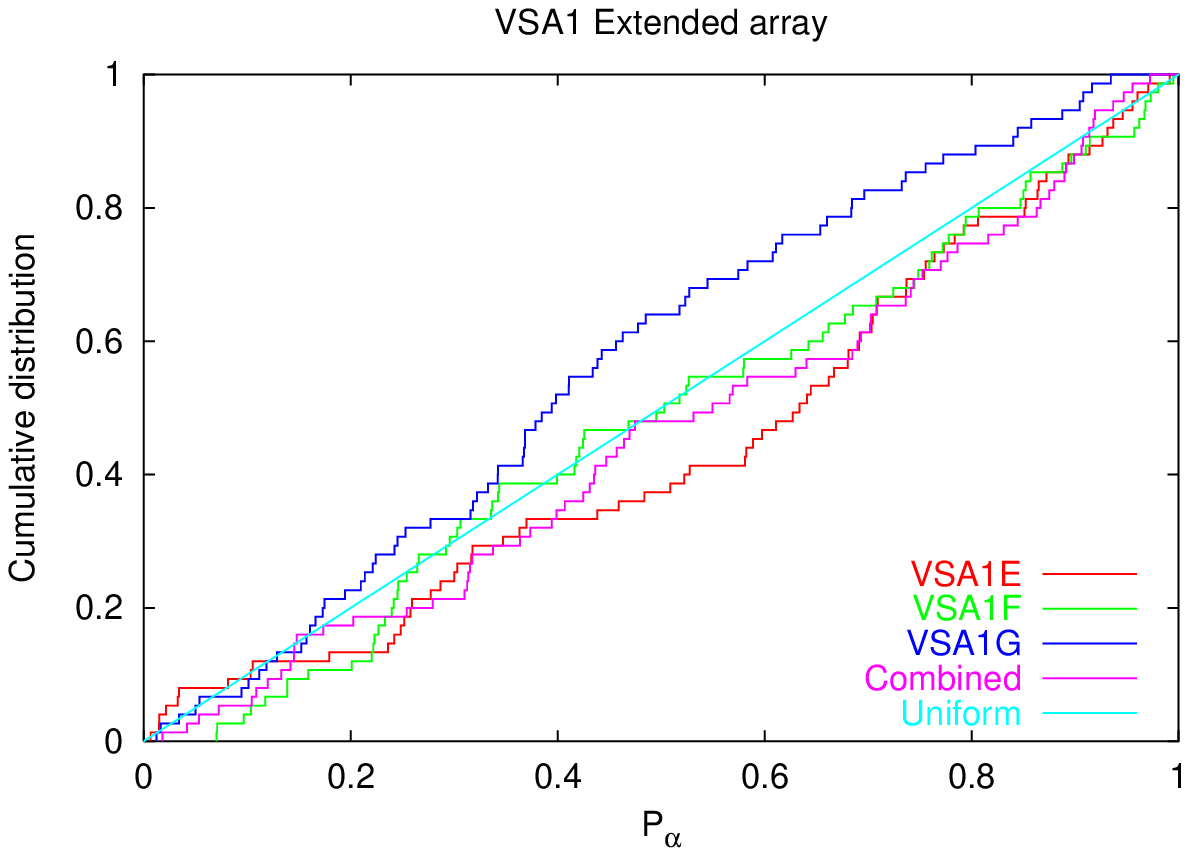}
\par}

\caption{\label{fig:p_alpha}Cumulative distribution of $P_{\alpha}$
for the VSA1 compact and extended arrays.}
\end{figure}

We wish to perform a quantitative comparison between our data and
the simulations, in order to test the null hypothesis, $H_{0}$,
that the data values were drawn from the same distribution as
the Gaussian simulations. Following the method in \citet{Komatsu_COBE},
for each bispectrum estimate $\hat B_{\alpha }$ we compute the 
quantity
\begin{equation}
P_{\alpha }
=\frac{N\left( \left| \widehat{B}^{\mathrm{SIM}}_{\alpha }\right| 
<\left| \widehat{B}^{\mathrm{VSA}}_{\alpha }\right
| \right) }
{N_{\mathrm{Total}}},
\end{equation}
where $\alpha$ represents a particular set of $\{\ell_{1},\ell_{2},\ell_{3}\}$
and $N_{\mathrm{Total}}=15000$, the total number of simulations. This
gives the probability that the magnitude of the bispectrum estimate
drawn under $H_0$ is less than the magnitude of the bispectrum
estimate obtained from the data. If $H_{0}$ is true, then the 
distribution of $P_{\alpha }$ is uniform on the interval [0,1]. 
If there is a tendency towards a non-zero bispectrum we could 
(na\"\i vely) expect to obtain more values of $P_{\alpha }$ close to 1.
Fig.~\ref{fig:p_alpha} shows the resulting cumulative distributions of
$P_{\alpha}$ obtained for the VSA1 observations, in comparison with the
straight line expected from a uniform distribution. 

The Kolmogorov-Smirnoff test operates by finding the quantity 
\begin{equation}
D=\mathrm{max}\left| F_{e}(P)-F(P)\right| ,
\end{equation}
where $F_{e}$ is the empirical distribution function defined by
\begin{equation}
F_{e}(P)=\frac{N(P_{\alpha }<P)}{N},
\end{equation}
and $F(P)$ is the continuous cumulative distribution function
for $P$ under $H_0$. If the $P_{\alpha}$ are independent,
identically distributed under $H_0$
then the distribution of $D$ is universal,
depending only on the number of values, $N$.

\begin{figure}
\includegraphics[height = 0.2\textheight]{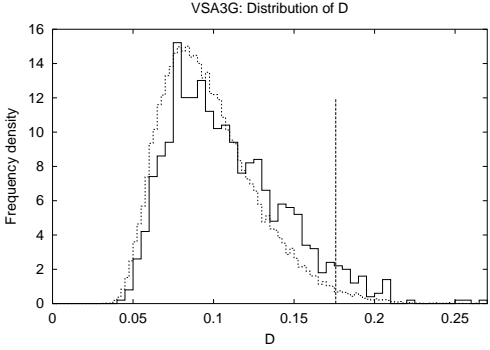}
\caption{\label{fig:ddist} The distribution of $D$ from simulations
(solid line) as compared with the distribution obtained if the $P_\alpha$
were independent (dotted line), together with the measured value from the
data for the field VSA3G.}
\end{figure}

Given $D=d$, one can then compute the tail probability $p$ which is given
by
\begin{equation}
p(d)=\mathrm{Prob}(D\geq d\mid H_{0}).
\end{equation}
In our case, we must calculate $p(d)$ using the distribution of $D$ obtained
from simulations, since the assumption that
the $P_{\alpha}$ are independent is invalid: although the formal covariance
between different bispectrum estimates is zero apart for the effects of the
primary beam, the same visibility point is used to calculate many different 
bispectrum estimates. Fig.~\ref{fig:ddist} shows the distribution
of $D$ obtained from simulations, compared with the distribution which
would be obtained if the $P_\alpha$ were independently drawn from a
uniform distribution. It can be seen that there is an increase of
larger values of $D$. A very similar distribution is 
obtained from noise-only simulations, when there is no correlation between
any of the visibility points.

Table \ref{table:p(d)} shows the calibrated results. It should be recognised
that this test gives us an indication of how well the simulations
match the data, which is dependent on a number of factors other than
simply whether the data are Gaussian: whether the power spectrum is accurate
and the correct telescope parameters have been used, and whether the
noise estimate is correct.
For comparison, when we compute the values of $P_{\alpha}$ by comparing
the VSA1F data with Gaussian simulations which have a beam of FWHM 4\fdg6 
rather than 2\fdg05, we obtain $p(d)=0.006$. Altering the amplitude of
the power spectrum by $\sim$10 per cent was found to change only slightly
the distribution of $d$, acting to alter $p(d)$ by $\sim$0.01.
\begin{table}
\caption{\label{table:p(d)}$d$-statistic and corresponding calibrated
tail probability for the various fields. The number of bispectrum
estimates for each field is $n$. The field labelled `VSA1 Compact'
represents the weighted average bispectrum from the individual
VSA1 compact fields, and likewise for the others.}
\begin{center}
\begin{tabular}{|l|c|c|c|}
\hline 
Field & $n$ & $d$ & $p(d)$ \\
\hline 
VSA1 & 171 &  0.076 & 0.38\\
VSA1A & 106 & 0.064 & 0.81 \\
VSA1B & 106 & 0.16 & 0.03 \\
VSA1 Compact & 172 & 0.086 & 0.25 \\
VSA1E & 75 & 0.17 & 0.06 \\
VSA1F & 75 & 0.10 & 0.52 \\
VSA1G & 75 & 0.15 & 0.11 \\
VSA1 Extended & 75 & 0.11 & 0.34 \\
VSA2 & 173 & 0.093 & 0.23 \\
VSA2-OFF & 174 & 0.084 & 0.29 \\
VSA2 Compact & 174 & 0.069 & 0.46 \\
VSA2E & 76 & 0.096 & 0.56\\
VSA2F & 76 & 0.070 & 0.87 \\
VSA2G & 95 & 0.107 & 0.43 \\
VSA2 Extended & 95 & 0.103 & 0.43 \\
VSA3 & 176 & 0.118 & 0.05\\
VSA3A & 106 & 0.068 & 0.76\\
VSA3B & 106 & 0.050 & 0.97\\
VSA3 Compact & 176 & 0.084 & 0.23\\
VSA3E & 97 & 0.107 & 0.45 \\
VSA3F & 76 & 0.090 & 0.62 \\
VSA3G & 76 & 0.176 & 0.05 \\
VSA3 Extended & 97 & 0.111 & 0.39 \\
\hline
\end{tabular}
\end{center}
\end{table}
Four out of the 17 individual fields have values of $p(d)$ which are
fairly low: VSA1B, VSA1E, VSA3 and VSA3G. For the total number of
fields analysed
we would expect an average of one to have $p(d) \leq 0.06$ by chance.
The lowest value of $p(d)$ is for the VSA1B data, at 3 per cent. This
field was observed for a total integration time of only 68 h, 
in comparison with around 200 hours for the majority of compact
array fields \citep{Taylor}, and consequently the mean error on
the visibilities is approximately twice that of the other fields.
Furthermore, the large value of $d$ is caused by an excess of low
values of $P_{\alpha}$, as can be seen in Fig.~\ref{fig:p_alpha}.
This indicates that we should be very hesitant about drawing any
conclusions about non-Gaussianity in the VSA1B field. The large 
value of $d$ for the VSA3G field is also due to an excess of low
values of $P_{\alpha}$, whereas there is an excess of high values
for the VSA3 and VSA1E fields. The VSA3 data have a greater extent
in the $uv$ plane than the other two VSA3 compact array fields (hence 
the high value of $n$). On eliminating all the bispectrum values
which are not calculated for the other two fields, we obtain
$n = 106$, $d = 0.149$ and $p(d) = 0.034$, so the significance level
is increased slightly, indicating that the large bispectrum values
are not concentrated in the region of large $\ell$. The VSA2 compact
array fields, which have the longest integration times, are
consistent with the simulations.

By testing the data in this way, we are making no assumptions about the
type of non-Gaussianity that we are looking for. This means that our test
is very general, but not very powerful as it is not tailored for optimal
detection of any particular non-Gaussian signature. One disadvantage of 
the test is that all of the $P_{\alpha}$ values are given equal consideration,
regardless of the fact that some bispectrum estimates are more dominated by
noise than others. In the following section we consider two tests which are
tailored for detecting particular types of non-Gaussianity: that arising
from point sources, and from the time of recombination.

\subsection{Point sources}
\label{sec:point}

\subsubsection{Theory}

A point source of strength $s$ at position $\bmath{x}_0$
can be described by
\begin{equation}
\Delta I(\bmath{\hat x}) 
= s \,\delta(\bmath{\hat x}-\bmath{\hat x}_0).
\end{equation}
The contribution this makes to the visibilities is
\begin{equation}
\Delta V(\bmath{u}) 
= s\, A(\bmath{\hat x}_0)\,
e^{2\pi i {\bmath{\hat x}_0}.\bmath{u}},
\end{equation}
and therefore the contribution to the bispectrum is
\begin{eqnarray}
\nonumber
\Delta V(\bmath{u}_1)\Delta V(\bmath{u}_2)
\Delta V(\bmath{u}_3) \!&
= \! & s^3 \, A^3(\bmath{\hat x}_0)\,
e^{2\pi i {\bmath{\hat x}_0}.
(\bmath{u}_1+\bmath{u}_2+\bmath{u}_3)}\\
& = & s^3 \, A^3(\bmath{\hat x}_0).
\label{eq:pointsource}
\end{eqnarray}
There will also be cross terms such as 
$V(\bmath{u}_1)V(\bmath{u}_2)\Delta V(\bmath{u}_3)$
but these will have a mean value of zero. In addition, if we have several
point sources there will be additional cross terms introduced. 
However, on average the contribution to the reduced bispectrum arising
from point sources should be uniform \citep{Komatsu_fnl}: 
\begin{equation}
B^{\mathrm{ps}}(\ell_{1},\ell_{2},\ell_{3})=B^{\mathrm{ps}}.
\end{equation}
So we can form an estimate of the point source contribution simply by calculating 
the (weighted) mean value of the bispectrum. 
\subsubsection{Point sources in the VSA data}
\begin{figure}

\includegraphics[width = \linewidth]{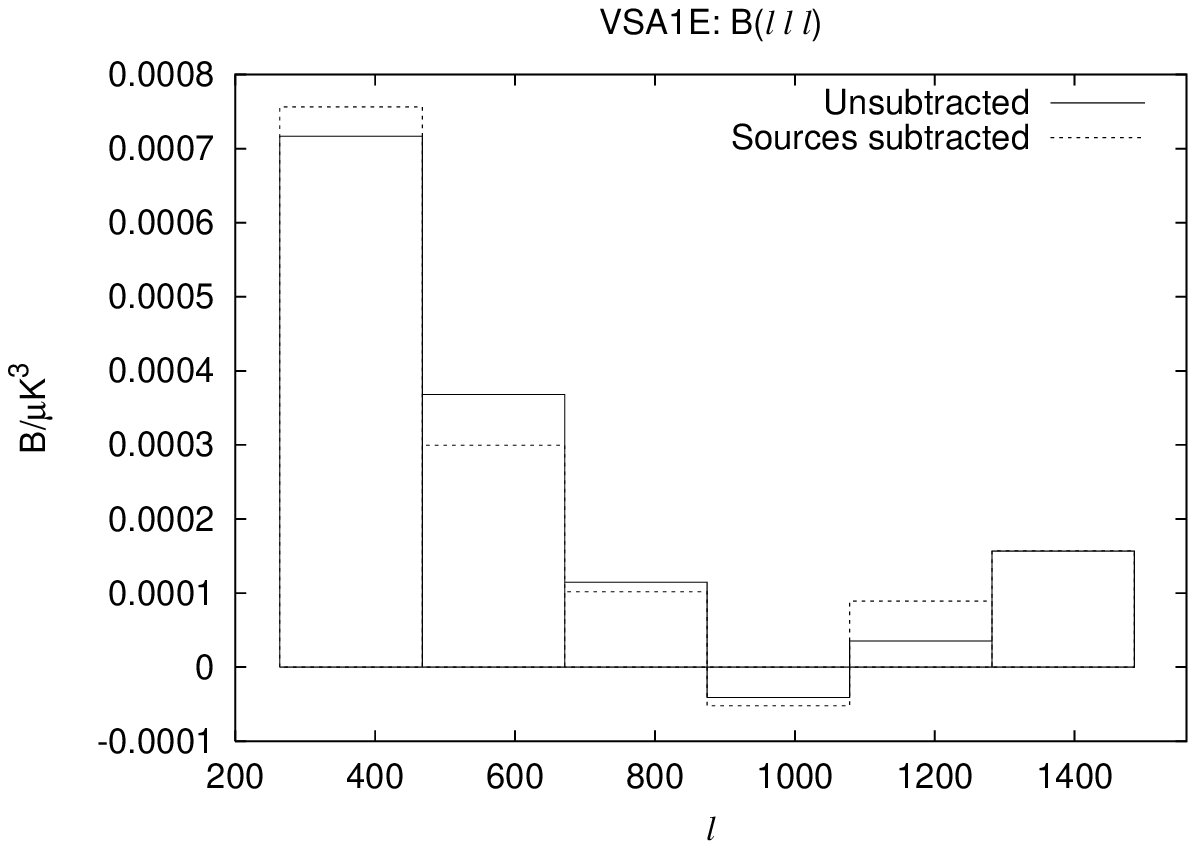}
\includegraphics[width = \linewidth]{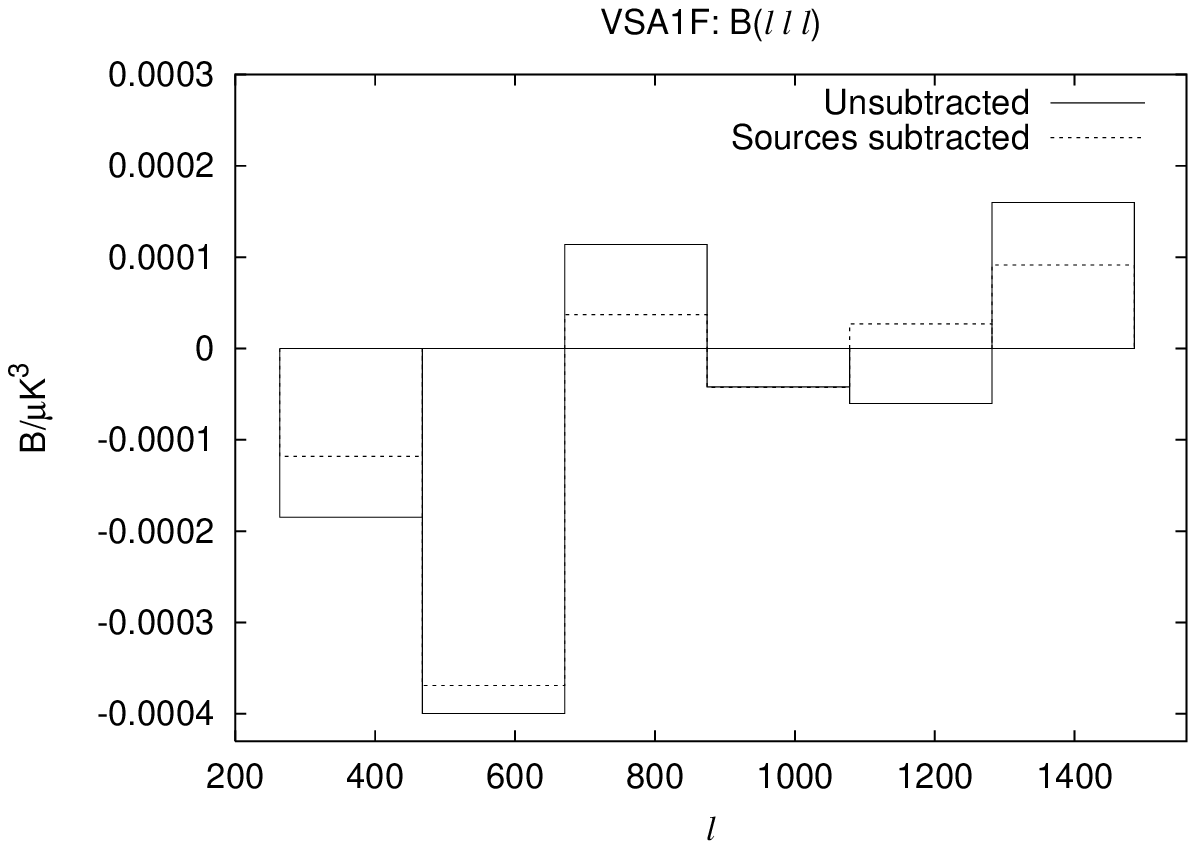}

\caption{Diagonal component of the bispectrum for the fields VSA1E and
VSA1F before and after point source subtraction, illustrating the 
level of change. \label{fig:source_sub}}

\end{figure}

We compared the estimated value of $B^{\mathrm{ps}}$ from the extended
array data before and after source subtraction \citep{Taylor, Grainge}.
The results are shown in Table \ref{table:meanbi}, and illustrated
in Fig.~\ref{fig:source_sub}.
\begin{table}
\caption {Mean bispectrum values before and after point source subtraction for
the three fields in the VSA1 region observed with the extended array.}
{\centering \begin{tabular}{|l|r|r|r|}
\hline
Field &
$\widehat{B}^{\mathrm{ps}} /(10^{-5} \mu \mathrm{K}^3)$ & 
$\widehat{B}^{\mathrm{ps}} /(10^{-5} \mu \mathrm{K}^3)$ &
$\sigma /(10^{-5} \mu \mathrm{K}^3)$\\
& Raw data & Sources subtracted & \\
\hline
VSA1E & 1.19 & 0.92 & 1.19 \\
VSA1F & -0.96 & -1.61 & 1.30 \\
VSA1G & 1.01 & 1.09 & 1.34 \\
\hline 
\end{tabular}\par}
\label{table:meanbi}
\end{table}
The mean value of the bispectrum is altered by 
$\sim$1--6$\times 10^{-6} \mu \mathrm{K}^3$ when the point sources are subtracted.
Therefore, although subtracting the sources
does change the bispectrum estimates, the resulting difference
is considerably smaller than the standard deviation on the mean value of the
bispectrum which arises from sample variance and noise, and so it does not seem
possible to detect point sources in this way. However, if future VSA
observations have a much lower noise and probe higher values of $\ell$
then it may be possible to use
the bispectrum to detect the presence of point sources.
\subsubsection{Residual point sources}
We can estimate the contribution to the bispectrum from residual point
sources using the results of the 15-GHz Ryle survey \citep{Waldram}. This
survey found that the differential source count at 15 GHz could be well
approximated by
\begin{equation}
n(S) \approx K \left ( \frac{S}{\mathrm{Jy}} \right )^{-\beta}
	 \mathrm{Jy^{-1} sr^{-1}},
\end{equation}
where $S$ is the flux density, $\beta = 2.15$ and $K = 51$. 
Assuming a spectral index
$\alpha = 0.55$ we find that, at 34 GHz, $K = 30$.

An individual point source contributes to the three-point visibility
function as described
by equation (\ref{eq:pointsource}). Assuming a Poisson distribution of
sources, the mean contribution from unsubtracted sources can be 
expressed as
\begin{eqnarray}
\langle V(\bmath{u}_1) V(\bmath{u}_2) V(\bmath{u}_3) \rangle & = &
\int A^3 (\bmath{\hat x}) \; d^2 \bmath{\hat x} 
\int_0^{S_*} S^3 n(S) \; dS \nonumber \\
& = & \frac{2}{3} \pi \sigma^2 K \frac{S_*^{4-\beta}}{4-\beta},
\end{eqnarray}
for $\bmath{u}_1+\bmath{u}_2+\bmath{u}_3 = 0$,
where $S_*$ is the source subtraction level. 

For the compact array, taking $S_*$ = 80 mJy we obtain
a theoretical contribution from unsubtracted sources of 
$\sim 4 \times 10^{-4} \mathrm{Jy^3}$ (equivalent to 
$4 \times 10^{-6} \mu\mathrm{K}^3$)
and for the extended array, taking $S_*$ = 20 mJy we obtain
$\sim 6 \times 10^{-6} \mathrm{Jy^3}$ (equivalent to 
$3 \times 10^{-7} \mu\mathrm{K}^3$).
Comparing this with the
error on the mean value of the bispectrum of $\sim 10^{-4}
\mu \mathrm{K^3}$ for the compact array and $\sim 10^{-5}
\mu \mathrm{K^3}$ for the extended array, we see that we do not 
expect the bispectrum to detect the presence of the residual
sources.

If we take these values and consider the comparison with the 
predicted bispectrum arising from the coupling between the
Sunyaev-Zel'dovich effect and weak lensing effects, as given by
\citet{Komatsu_fnl}, we find that the SZ-lensing bispectrum
will be overwhelmed by residual point sources at the source
subtraction level of the extended array for $\ell \gtrsim 200$.
At lower values of $\ell$ we estimate that the SZ-lensing
contribution to the bispectrum is approximately four orders of
magnitude smaller than the error on our bispectrum estimates.
Therefore our data is not sensitive to this effect.

\subsection{Simulated point sources}
\label{sec:simp}

\subsubsection{Noise-only simulations}

As a preliminary test we subtracted a single point source from an
extended-array simulation
with no noise or CMB signal. The resulting mean bispectrum%
\footnote{By `bispectrum' in this section we mean the value of 
the three-point function as given by equation (\ref{eq:3point})
since it is natural to work in units of Jy when considering point
sources.} was
96 per cent of the theoretical value of $s^3 A^3(\bmath{\hat x}_0)$,
with all the individual bispectrum estimates being very similar in value.
The discrepancy is due to the fact that we slightly relax the requirement
that $\bmath{u}_1+\bmath{u}_2+\bmath{u}_3 = 0$
and so the phase in equation (\ref{eq:pointsource}) is slightly non-zero.

On subtracting two point sources (of flux density 1.0 Jy and 0.89 Jy after
attenuation) from the same simulation, we immediately
find that each individual bispectrum estimate is very different, varying
from -0.048 Jy$^3$ to -2.4 Jy$^3$, as a result of the cross-terms. The mean
bispectrum value is -1.3 Jy$^3$ compared to a theoretical value of -1.7 Jy$^3$.
On subtracting the VSA1E source list, 10 sources in total, the mean bispectrum
value is -2.95$\times 10^{-5} \mathrm{Jy}^3$  in comparison with the
theoretical value of -3.53$\times 10^{-5} \mathrm{Jy}^3$,
and there is less variation between individual bispectrum estimates, suggesting
that with more sources the cross-terms have a greater tendency to cancel.

In order to ascertain the effect of noise on our ability to detect 
point sources we used sets of simulated data
with no CMB component, just noise, to which we added the point sources
subtracted from the VSA1E field. Each simulation had exactly the
same $uv$ and noise template apart from an overall scaling factor on the
noises. The values of the mean bispectra, together with the standard
deviation of the mean bispectrum obtained from noise-only simulations,
are shown in Table \ref{table:sources}. 
\begin{table}
\caption{\label{table:sources} The rms noise, mean bispectrum values
and standard deviations for noise-only simulations with point sources added.}
{\centering \begin{tabular}{|l|c|c|}
\hline
rms noise / Jy &
$\langle B \rangle /10^{-5}\mbox{Jy}^3 $ &
$\sigma /10^{-5}\mbox{Jy}^3$ \\
& [/$10^{-5}\mu \mathrm{K}^3$] & [/$10^{-5}\mu \mathrm{K}^3$] \\
\hline 
2.2   &  8.8 [4.4]  & 390 [19] \\
0.44  & 3.59 [0.18] & 3.1 [0.15] \\
0.22  & 1.89 [0.093]& 0.39 [0.019]\\
0.044 & 2.13 [0.11] & 3.1$\times 10^{-3}$ [1.5$\times 10^{-4}$]\\
0.0044& 2.22 [0.11] & 3.1$\times 10^{-6}$ [1.5$\times 10^{-7}$]\\
\hline
\end{tabular}\par}
\end{table}
\begin{figure}
\centering
\includegraphics[height = 0.2\textheight]{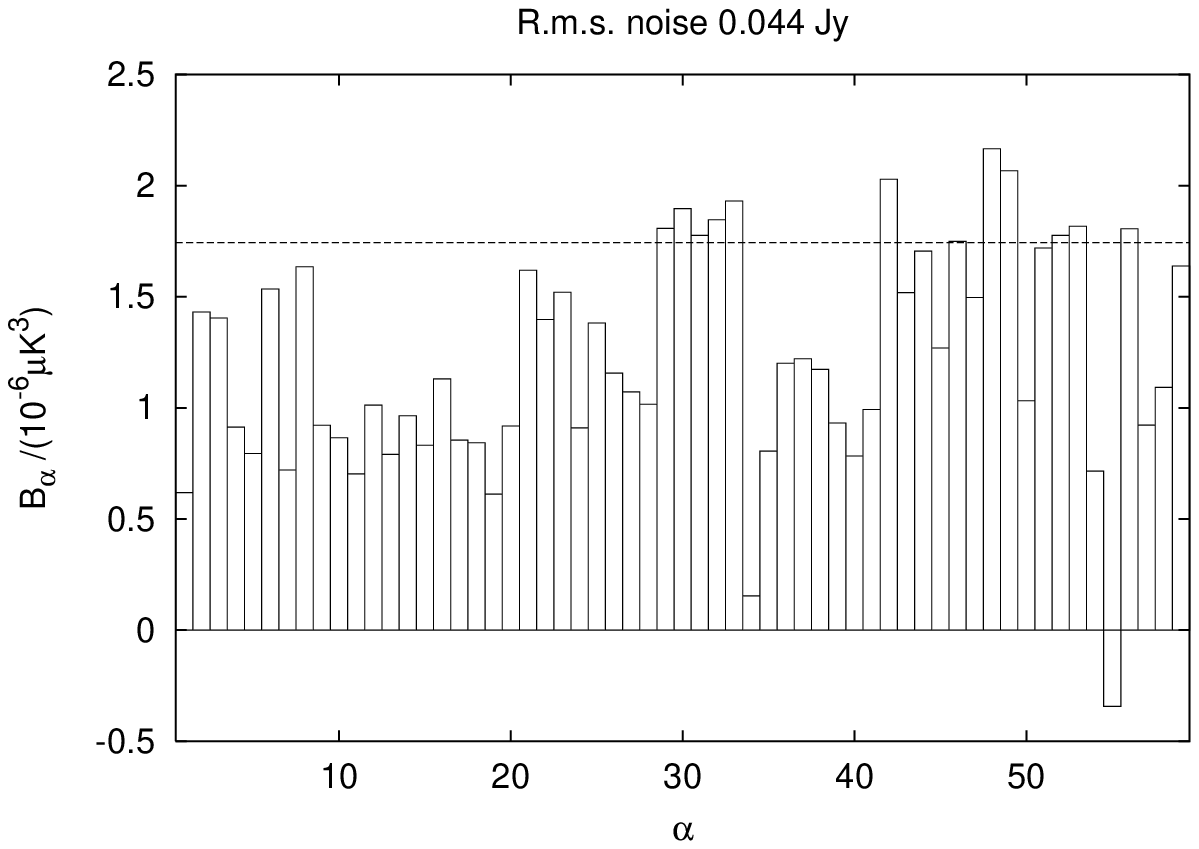}
\includegraphics[height = 0.2\textheight]{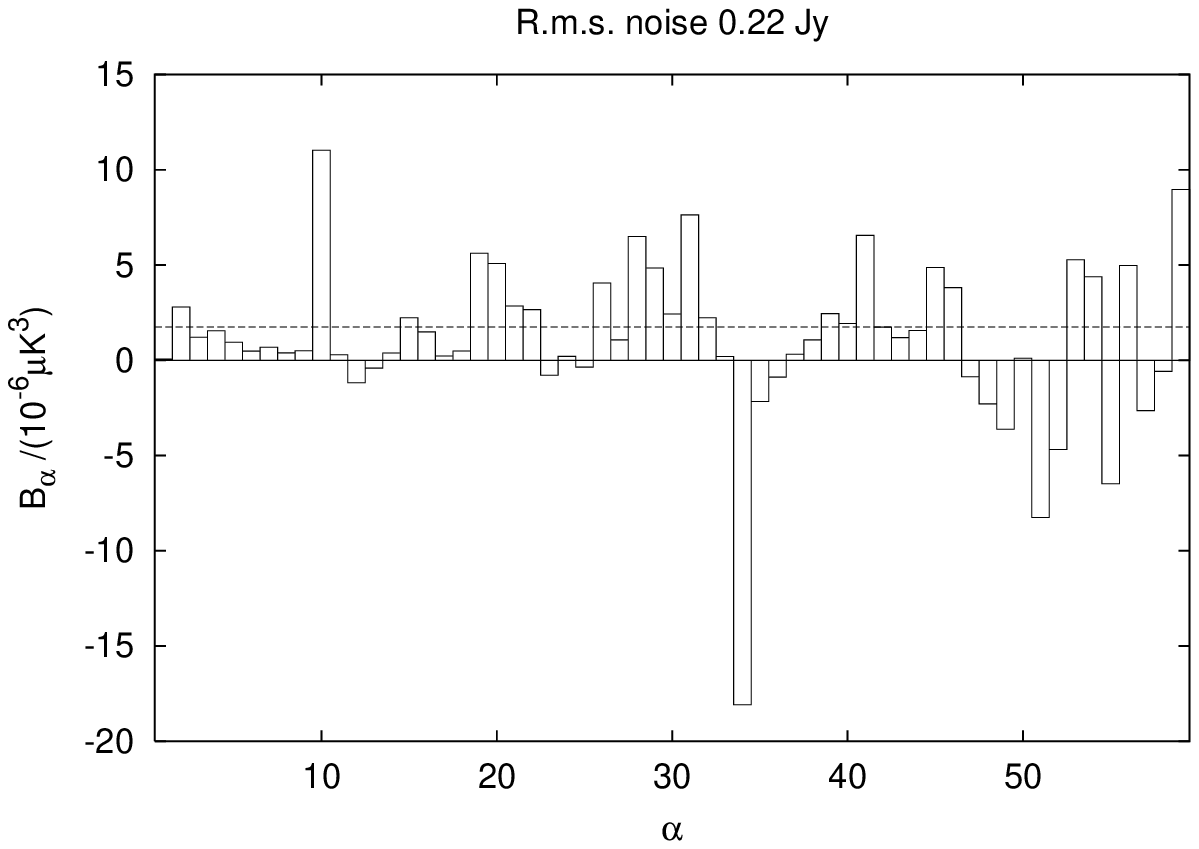}
\includegraphics[height = 0.2\textheight]{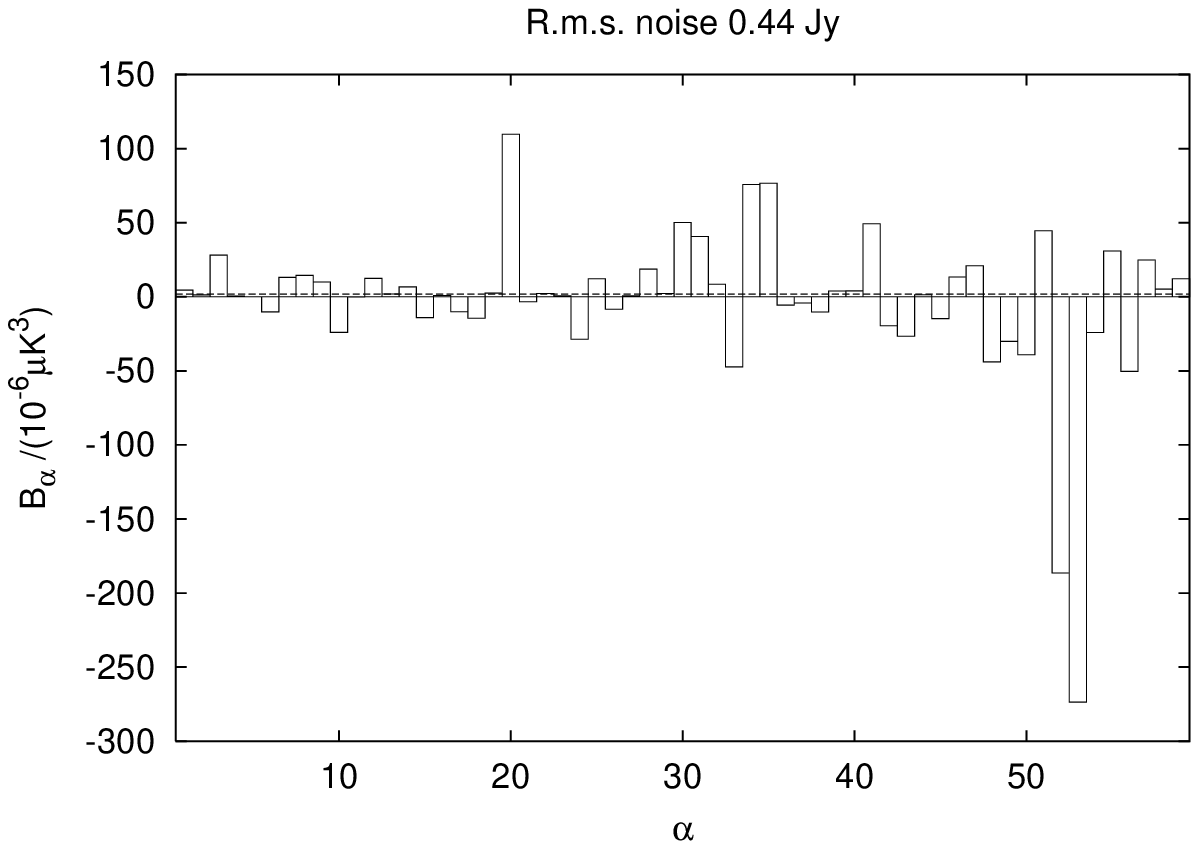}

\caption{\label{fig:source} Bispectrum estimates from 
noise-only simulations with point sources added, with theoretical
bispectrum plotted as a horizontal line. $\alpha$ represents
a particular set of $\{\ell_1, \ell_2, \ell_3\}$}

\end{figure}
The mean value of the bispectrum is positive in all cases, but a definite
detection of point sources is obtained only when the rms noise is less 
than 0.44 Jy. The point source bispectrum is 
completely swamped by noise for the case with the highest noise.
Fig.~\ref{fig:source}
illustrates the bispectrum components for each case.

There were 10 sources in the simulations in total, and the sum 
$\sum_{i=1}^{10} A^3(\bmath{\hat x}_i) s_i^3$ is 
$3.5 \times 10^{-5} \mbox{Jy}^3$. This is 1.6 times the 
mean bispectrum value calculated, weighting according to the
variance from simulations. However, if we use an alternative weighting scheme, 
according to the number of triangles of vectors which are combined to form each
bispectrum estimate, we obtain (for the lowest noise case) a mean bispectrum
value of $2.6 \times 10^{-5} \mbox{Jy}^3$. The cross-terms introduce a lot of
variation between the different individual bispectrum estimates and hence cause
the final values to depend on the weighting scheme. 

The rms noise on each visibility point for the VSA1E data is 
$\sim$1 Jy, which is at a level at which the contribution from point sources
is swamped by noise. In addition, the CMB itself will increase further the 
variance on the mean value of the bispectrum.
Therefore we cannot
expect that we will detect the presence of sources in the extended 
array data using this method.

For comparison, we also performed the Kolmogorov-Smirnoff test on the
point source simulations. This gave a conclusive detection of non-Gaussianity
for the cases with a rms of 0.0044 Jy and 0.044 Jy but no detection
when the rms noise was 0.22 Jy [the value of $p(d)$ was 0.42] despite the
conclusive detection found by computing the mean value of the bispectrum.
This illustrates how using a tailored statistic when searching for a particular
non-Gaussian signal is a much more powerful method of detection than using a
general statistic. 

\subsubsection{CMB simulation with strong sources}

\begin{figure}
\centering
\includegraphics[height = 0.3\textheight,angle = -90]{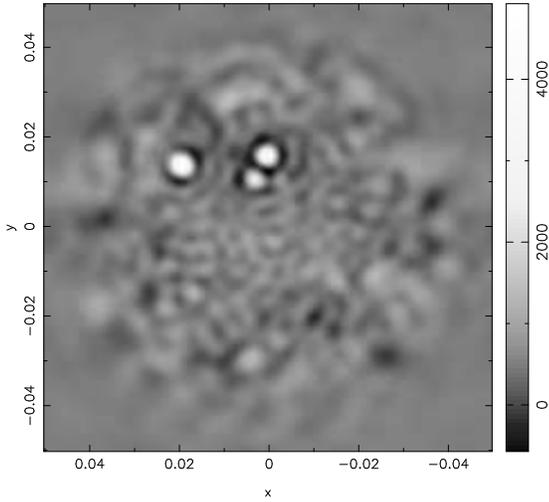}
\caption{\label{fig:strong_sources}Simulated CMB with point sources, convolved
with the primary beam.}
\end{figure}

We computed the mean bispectrum from a simulation based on the noise
and $uv$-positions of an extended-array field, with both CMB and 
bright sources. The observed sky is shown
in Fig.~\ref{fig:strong_sources} with the point sources clearly visible
in the map. The mean bispectrum value was $0.3$ Jy$^3$, a factor
of 1000 greater than the standard deviation from simulations with no
point sources.
The bispectrum is sensitive to 
the point sources as (source strength)$^3$ and so is useful for
detecting strong sources but not for weak sources.

\subsection{Primordial Non-Gaussianity}
\label{sec:fnl}

It has become usual to quantify the primordial non-Gaussianity
which arises from weak non-linear evolution on super-Hubble scales
by a single parameter $f_{\mathrm{NL}}$, the non-linear coupling parameter:
\begin{equation}
\mathcal{R}(\bmath{x}) = \mathcal{R}_G(\bmath{x})
	+ f_{\mathrm{NL}}(\mathcal{R}_G^2(\bmath{x}) 
 	- \langle \mathcal{R}_G^2(\bmath{x}) \rangle),
\end{equation}
where $\mathcal{R}(\bmath{x})$ is the comoving curvature perturbation 
as defined according to the convention in \citet{Liddle},
and $\mathcal{R}_G(\bmath{x})$ its Gaussian component\footnote{Some
authors \citep{Komatsu_fnl,Santos} use a slightly different definition 
of $f_{\mathrm{NL}}$ based on 
$\Phi(\bmath{x}) = \frac{2}{3}\mathcal{R}(\bmath{x})$ for adiabatic
perturbations in radiation domination. This ignores neutrino anisotropic
stress which produces a five per cent correction.} \citep{Acquaviva,Maldacena}.
Even if the fluctuations produced by inflation are perfectly Gaussian, 
a non-zero $f_{\mathrm{NL}}$ will be at second-order in perturbation 
theory due to the non-linear nature of general relativity \citep{Bartolo}.
The current distribution of matter, which is highly non-Gaussian,
is a result of the non-perturbative non-linear coupling
between modes, which becomes increasingly significant after
recombination as the perturbations grow. Previous studies have not
detected any significant primordial non-Gaussianity, but have simply
put upper limits on it. This is not surprising as the predictions from
all but the most contrived inflationary models are well below the
current upper limits. The recent results from {\sl WMAP} 
place an upper limit of 89 on the value of $f_{\mathrm{NL}}$ \citep{WMAP},
in comparison with predictions from slow-roll inflation which give
$ \left| f_{\mathrm{NL}}\right| \sim 10^{-1}$--$10^{-2}$, apart from
the non-Gaussianity introduced by the subsequent evolution of the perturbations.

We should therefore only expect to place limits on the 
value of $f_{\mathrm{NL}}$ using the VSA data, but it is interesting to
see what limits the data are capable of producing. In addition, the
techniques developed to estimate $f_{\mathrm{NL}}$ can easily be applied
to any other theoretical bispectrum which has its amplitude as 
the only free parameter.

\begin{figure}
\centering
\includegraphics[height = 0.24\textheight]{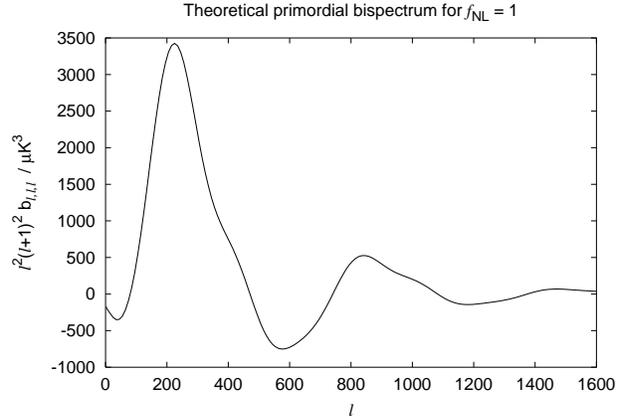}
\caption{\label{fig:fnlraw}Diagonal component of the theoretical bispectrum
for $f_{\mathrm{NL}} = 1$.}
\end{figure}

We calculated the theoretical bispectrum for the case 
$f_{\mathrm{NL}}$=1 using a modified version of CAMB \citep*{Lewis}.
The resulting bispectrum has features on the usual acoustic scale
of the power spectrum, and its diagonal component is shown in
Fig.~\ref{fig:fnlraw}.

\begin{figure}
\centering
\includegraphics[height = 0.24\textheight]{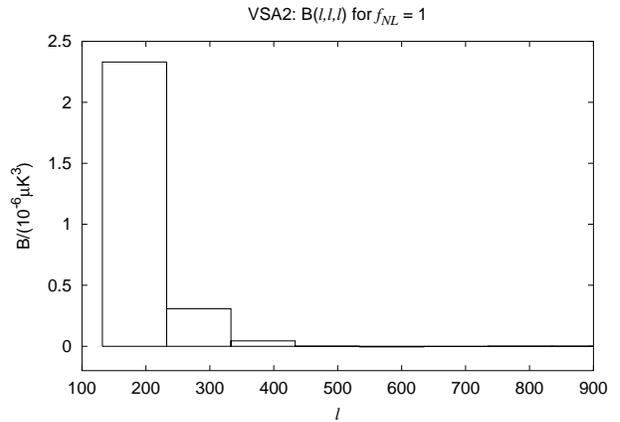}
\caption{\label{fig:fnlVSA}Diagonal component of theoretical measured
bispectrum for the VSA2 data with $f_{\mathrm{NL}} = 1$}
\end{figure}

In order to relate the theoretical bispectrum to the measured three-point
visibility function, we performed the integral described in 
Appendix~\ref{sec:appc} which convolves the bispectrum with the
aperture function. For each field, we
then estimate the set of $Q_\alpha$, the theoretical values of the
$B_\alpha$ for $f_{\mathrm{NL}}=1$ [where $\alpha = (\ell_1, \ell_2, \ell_3)$]
according to equation (\ref{eq:estimator}) using the predicted value 
of $\langle S(\bmath{u}_1) S(\bmath{u}_2) S(\bmath{u}_3) \rangle$
as calculated from the integral. Figure~\ref{fig:fnlVSA} shows
the resulting diagonal component of the theoretical measured bispectrum.

We estimate the value of $f_{\mathrm{NL}}$ from our bispectrum 
estimates with the estimator
\begin{equation}
\widehat f_{\mathrm{NL}} = \frac
	{\sum_\alpha \frac{\hat B_\alpha Q_\alpha}{\sigma^2_\alpha}}
	{\sum_\beta  \frac{Q^2_\beta}{\sigma^2_\beta}},
\label{eq:fnl_est}
\end{equation}
where the $\sigma^2_\alpha$ are the variances of the bispectrum estimates
from Gaussian simulations. We obtain these variances from mosaiced
simulations to allow us properly to include mosaiced fields. Equation
(\ref{eq:fnl_est}) reduces to the optimal linear estimator of
$f_{\mathrm{NL}}$ from the bispectrum estimates in the limit that
these are uncorrelated.

The overall estimate of $f_{\mathrm{NL}}$ obtained from the compact array data
is 85, with a standard deviation of 2700. For the extended array, 
we obtain an estimate of -400 with a standard deviation of 3500.
The 95 per cent confidence limits are 5400 and 7000 for the compact and
extended arrays respectively. 
These contraints are slightly weaker than those obtained from the 
{\sl MAXIMA} data using the FFT estimator \citep{Santos}, due probably to
the fact that the {\sl MAXIMA} data extend to lower multipoles with slightly
higher $\Delta \ell$ resolution, although the VSA sky coverage is slightly
greater. [The VSA compact array data cover 101 deg$^2$ \citep{Taylor}; the 
{\sl MAXIMA} data used had an area of 60 deg$^2$.] This is also
why the compact array data are better at constraining $f_{\mathrm{NL}}$
than the extended array data. The bispectrum arising from primordial
non-Gaussianity falls quickly with $\ell$ and so extending the measurements
to higher values of $\ell$ does not significantly improve the constraints
(unless the noise is very low). This is why the {\sl WMAP} data, which are 
cosmic-variance limited at low $\ell$, are able to place so much tighter
constraints on $f_{\mathrm{NL}}$.


\section{Conclusions}

Our bispectrum calculations indicate a slightly greater discrepancy
with the Gaussian simulations for the fields VSA1B, VSA1E, VSA3,
and VSA3G than we would expect if the CMB
sky were perfectly Gaussian. However, there is little discernible
pattern in the way in which the data deviate from the simulations.
A small level of non-Gaussianity is to be expected as there will
inevitably be a degree of foreground contamination and unsubtracted
point sources, but we have shown that in the case of point sources
we do not anticipate to be able to detect the resulting non-Gaussianity
with the current level of experimental noise. In the case
of point sources, the CMB itself acts as a level of noise that can
only be reduced by taking measurements at higher $\ell$.
We note that \citet{Savage} found some evidence for non-Gaussianity
in the VSA1 mosaic, which was attributed to point sources or contamination
from galactic foregrounds.
It is possible that the excess of large bispectrum values
in the VSA1E field arises from the same cause. The fact the low values
of the tail probability $p(d)$ in a Kolmogorov-Smirnoff test 
only appear in isolated individual fields indicate that, if
the non-Gaussianity which is hinted at is real, it is localised in space.

The limit on the largest scales probed by the VSA means that it can only
be used to place weak constraints on the value of the quadratic
non-Gaussianity parameter $f_{\mathrm{NL}}$. The
data are more suited to detecting non-Gaussianity that is present
on smaller scales.


\section*{Acknowledgements}
We thank the staff of the Mullard Radio Astronomy Observatory,
the Jodrell Bank Observatory and the Teide Observatory for invaluable
assistance in the commissioning and operation of the VSA. The VSA is
supported by PPARC and the IAC.
AC acknowledges a Royal Society University Research Fellowship.
GR acknowledges a Leverhulme Fellowship at the University of Cambridge.
RSS, SS, NR and KL acknowledge support by PPARC studentships.
CD thanks PPARC for funding a post-doctoral research associate position
for part of the work. We thank J. Magueijo for useful discussions, and
M. Kuntz for valuable comparisons on material presented in Section 4.6.


\bibliography{ng_refs,cmb_refs}
\bibliographystyle{mn2e}


\appendix
\onecolumn
\section{Optimal cubic bispectrum estimators}
\label{sec:appa}

We have estimated the bispectrum using a simple cubic estimator with each
visibility weighted by the inverse of its variance
(see equation~\ref{eq:estimator}).
This estimator is not optimal, but for interferometer data, where the
signal and noise covariances are nearly diagonal, it is close to optimal
as we shall argue in this appendix.

We begin by reviewing the calculation
of the optimal cubic estimator, first given by~\citet{Heavens} and discussed
further by \citet{Santos}. For simplicity, consider real data $\{ a_i \}$
whose three-point function is related to the bispectrum $B_\alpha$ that
we wish to estimate by
\begin{equation}
\langle a_i a_j a_k \rangle = \sum_\alpha Q^\alpha_{ijk} B_\alpha.
\label{eq:a1}
\end{equation}
Here, $\alpha$ denotes an (ordered) triplet $\ell_1 \geq \ell_2 \geq \ell_3$
and $Q^\alpha_{ijk}$ is totally symmetric in $i$, $j$ and $k$. We
construct an estimator $\hat{y}_\alpha$ which is cubic in the $\{a_i\}$,
\begin{equation}
\hat{y}_\alpha = \sum_{ijk} E^\alpha_{ijk} a_i a_j a_k,
\label{eq:a2}
\end{equation}
and which is unbiased,
\begin{equation}
\langle \hat{y}_\alpha \rangle = \sum_{ijk} E^\alpha_{ijk} \sum_\beta
Q^\beta_{ijk} B_\beta \quad \Rightarrow \quad \sum_{ijk} E^\alpha_{ijk}
Q^\beta_{ijk} = \delta_{\alpha \beta}. 
\label{eq:a3}
\end{equation}
Only the symmeterised part of $E^\alpha_{ijk}$ enters the estimator
$\hat{y}_\alpha$ but we do not impose total symmetry at this stage
as enforcing this complicates the variational procedure that follows.
Following~\citet{Heavens}, we construct the variance of $\hat{y}_\alpha$
under the assumption that the data are Gaussian (so the estimator is
optimised for only weakly non-Gaussian data), to find
\begin{equation}
\mathrm{var}(\hat{y}_\alpha) = \sum_{ijk} \sum_{i'j'k'}
E^\alpha_{ijk} E^{\alpha}_{i' j' k'}(C_{ii'}C_{jj'}C_{kk'} +
\mathrm{perms}),
\label{eq:a4}
\end{equation}
where $C_{ij} \equiv \langle a_i a_j \rangle$ is the (symmetric)
covariance of the data. Unlike~\citet{Heavens}, we have not imposed
symmetry of $E^\alpha_{ijk}$, so we cannot simplify equation~(\ref{eq:a4})
in the manner that he does (see his equation 16). We now vary $E^\alpha_{ijk}$
to minimise the variance of $\hat{y}_\alpha$, enforcing the constraints
on the mean (equation~\ref{eq:a3}) with a set of Lagrange multipliers
$\lambda_{\alpha \beta}$, to find
\begin{equation}
6 \sum_{ijk} E^\alpha_{(ijk)}(2 C_{ii'} C_{jj'} C_{kk'}
+ 3 C_{ij} C_{k(i'}C_{j'k')}) - \sum_\beta \lambda_{\alpha \beta}
Q^\beta_{i' j' k'} = 0.
\label{eq:a5}
\end{equation}
Note that only the symmetric part $E^\alpha_{(ijk)}$ enters this expression
and so only that part is constrained as expected. We can now solve
for a symmetric $E^\alpha_{ijk}$ and $\lambda_{\alpha \beta}$ to find
\begin{equation}
E^\alpha_{ijk} = \sum_\beta \lambda_{\alpha \beta} \sum_{i' j' k'}
Q^\beta_{i' j' k'} \left(\frac{1}{12} C^{-1}_{ii'} C^{-1}_{jj'}C^{-1}_{kk'}
- \frac{1}{4(4+N)} C^{-1}_{i' j'} C^{-1}_{k' (i} C^{-1}_{jk)} \right) 
\label{eq:a6}
\end{equation}
where
\begin{equation}
\lambda^{-1}_{\alpha \beta} = \frac{1}{12} \sum_{ijk} \sum_{i'j'k'}
Q^\alpha_{ijk} Q^\beta_{i'j'k'} \left(C^{-1}_{ii'} C^{-1}_{jj'}C^{-1}_{kk'} 
- \frac{3}{4+N} C^{-1}_{i' j'} C^{-1}_{k' i} C^{-1}_{jk} \right).
\label{eq:a7}
\end{equation}
In these expressions, $N$ is the number of data points and is typically
very large. Our equations~(\ref{eq:a6}) and (\ref{eq:a7})
correct the results given as
equations (21) and (25) in~\citet{Heavens}. The error in the latter analysis
arises because symmetry of $E^\alpha_{ijk}$ is assumed prior to the
variation, but is then not enforced during it. However, the terms in
error in~\citet{Heavens} are suppressed by $1/N$ for large $N$, and so
the differences from the results derived here are small for $N \gg 1$.

For interferometer data, the covariance matrix $C_{ij}$ is close to diagonal
with correlations limited roughly to the extent of the aperture function.
Multiplication by $C^{-1}_{ij}$ reduces to inverse variance weighting the
visibilities if the effect of the primary beam is neglected. Furthermore,
in this limit $Q^\alpha_{ijk}$ enforces $\bmath{u}_i + \bmath{u}_j
+ \bmath{u}_k = 0$, so that equation~(\ref{eq:a6}), with $N \gg 1$, reduces to
the heuristically-weighted estimator of equation~(\ref{eq:estimator}).
In practice,
the finite extent of the primary beam makes the heuristic estimator
somewhat sub-optimal. The necessary developments to compute the optimal
estimator are given in Appendix~\ref{sec:appc}, where we construct
$Q^\alpha_{ijk}$. However, given the need to perform large-volume Monte-Carlo
simulations to assess properly the significance of our results, employing
the optimal estimator would have been prohibitively slow for the analysis in
this paper.


\section{Interferometer measurements}
\label{sec:appb}

The actual values recorded by the VSA are flux density measurements in
Janskys. The receivers are calibrated primarily by observations of
Jupiter \citep{Taylor}, with the primary beam $A(\bmath{x})$ normalised
so that $A(0)=1$. The intensity fluctuations of the CMB can be connected
to the temperature fluctuations by
\begin{equation}
\Delta I(\bmath{\hat{x}},\nu) \approx 
\frac{\partial B(\nu,T)}{\partial T} \Big\vert _{T=T_0}
\Delta T_{\mathrm{cmb}}(\bmath{\hat{x}}).
\end{equation}
The visibility seen by the interferometer can be expressed as
\begin{equation}
S(\bmath{u})=
\int \Delta I(\bmath{\hat{x}})\;
A(\bmath{\hat{x}}) \:
e^{2\pi i\bmath{u.\hat{x}}}\: \mathrm{d^2} \bmath{\hat{x}}.
\end{equation}
The measured visibility $V(\bmath{\hat{u}})$ is related to this by
\begin{equation}
V(\bmath{u})=S(\bmath{u})+N(\bmath{u}),
\end{equation}
where $N(\bmath{u})$ is the noise on baseline $\bmath{u}$.
We can relate this to the values of $a(\bmath{u})$ by
\begin{eqnarray}
S(\bmath{u}) & = &
\frac{\partial B(\nu,T)}{\partial T} \Big\vert _{T=T_0} 
T_0 \ a(\bmath{u}) \star \tilde{A}(\bmath{u}) \\
& = & f a(\bmath{u}) \star \tilde{A}(\bmath{u})
\label{eq:v(u)}
\end{eqnarray}
where $f=94\times10^6 \mathrm{Jy \: sr^{-1}}$ for $\nu$ = 34 GHz and
a CMB temperature $T_0$ of 2.726K \citep{Mather}.
If we consider the variance of the measured signal we obtain
\begin{eqnarray*}
\left \langle V(\bmath{u})V^{\ast}(\bmath{u}) \right \rangle & = &
f^2
\int \! \mathrm{d^2}\bmath{u}_1
\int \! \mathrm{d^2}\bmath{u}_2\:
\tilde A(\bmath{u}\!-\!\bmath{u}_1)
\tilde A^\ast(\bmath{u}\!-\!\bmath{u}_2)
\left \langle a(\bmath{u}_1)a^\ast(\bmath{u}_2) \right \rangle
+ \sigma^2_{\bmath u} \\
& = &
f^2
\int \! \mathrm{d^2}\bmath{u}_1
\int \! \mathrm{d^2}\bmath{u}_2\:
C(u_1)
\delta(\bmath{u}_1-\bmath{u}_2)
\tilde A(\bmath{u}\!-\!\bmath{u}_1)
\tilde A^\ast(\bmath{u}\!-\!\bmath{u}_2)
+ \sigma^2_{\bmath u} \\
& \approx &
f^2 C(u)
\int \mathrm{d}^2 \bmath{u}_1 \left|A(\bmath{u}_1)\right|^2
+ \sigma^2_{\bmath u},
\end{eqnarray*}
if we make the approximation that $C(u)$ is constant
over the width of the aperture function.
(We use $u_i$ to denote $|\bmath{u}_i|$.) 
Here, $\sigma^2_{\bmath u}$ is the variance of the noise, 
which is statistically
independent of the visibility measurement.

If $A(\bmath{x}) = \exp({-|\bmath{x}|^2/2\sigma^2})$
then we find that 
\begin{equation}
\label{eq:variance}
\left \langle V(\bmath{u})V^{\ast}(\bmath{u}) \right \rangle
\approx \pi \sigma^2 f^2 C(u) + \sigma^2_{\bmath u}.
\end{equation}
Similarly, if we consider
\begin{eqnarray}
\left \langle V(\bmath{u}_1)V(\bmath{u}_2) V(\bmath{u}_3) \right \rangle
& = & f^3 \int \! \mathrm{d^2} \bmath{u}_1'
\int \! \mathrm{d^2} \bmath{u}_2' \int \! \mathrm{d^2} \bmath{u}_3'
\tilde A(\bmath{u}_1\!-\!\bmath{u}_1') 
\tilde A(\bmath{u}_2\!-\!\bmath{u}_2')
\tilde A(\bmath{u}_3\!-\!\bmath{u}_3')
B(u_1',u_2',u_3')
\delta^2(\bmath{u}_1'+\bmath{u}_2'+\bmath{u}_3')
\nonumber \\
\nonumber& \approx &
f^3 B(u_1,u_2,u_3)
\int \! \mathrm{d^2} \bmath{\hat x}
\int \! \mathrm{d^2} \bmath{u}_1'
\int \! \mathrm{d^2} \bmath{u}_2' \int \! \mathrm{d^2} \bmath{u}_3' 
\\ \nonumber & & \times
\exp[2\pi i \bmath{\hat x}.(\bmath{u}_1'+\bmath{u}_2'+\bmath{u}_3')]
\tilde A(\bmath{u}_1\!-\!\bmath{u}_1')
\tilde A(\bmath{u}_2\!-\!\bmath{u}_2')
\tilde A(\bmath{u}_3\!-\!\bmath{u}_3')\\
& = & f^3 B(u_1,u_2,u_3)
\int \! \mathrm{d^2} \bmath{x}  A^3(\bmath{\hat x})
\exp[2\pi i \bmath{\hat x}.(\bmath{u}_1+\bmath{u}_2+\bmath{u}_3)],
\label{eq:approxthreepoint}
\end{eqnarray}
assuming that the bispectrum varies only slowly over the width 
of the aperture function, and that the noise is Gaussian.
Taking the above form for $A(\bmath{x})$ and the
case $\bmath{u}_1+\bmath{u}_2+\bmath{u}_3=0$
we obtain
\begin{equation}
\left \langle V(\bmath{u}_1)V(\bmath{u}_2)
              V(\bmath{u}_3) \right \rangle
\approx f^3 \frac{2}{3}\pi \sigma^2 B(u_1,u_2,u_3).
\end{equation}

\section{Visibility three-point function}
\label{sec:appc}

In this appendix we derive an integral expression for the three-point
function of the interferometer visibilities. For the special case of
a Gaussian beam, we are able to reduce the integral further to obtain
a closed-form expression for the quantity $Q^\alpha_{ijk}$ introduced
in Appendix~\ref{sec:appa}.

Our starting point is the first equality in
equation~(\ref{eq:approxthreepoint}) which expresses the three-point
function of the visibilities as an integral over the bispectrum:
\begin{equation}
\langle S(\bmath{u}_1) S(\bmath{u}_2) S(\bmath{u}_3) \rangle =
\int_{-\infty}^\infty \rmn{d}^2 \bmath{u}_1'
\int_{-\infty}^\infty \rmn{d}^2 \bmath{u}_2'
\int_{-\infty}^\infty \rmn{d}^2 \bmath{u}_3'
B(u_1', u_2', u_3') \tilde{A}(\bmath{u}_1 - \bmath{u}_1')
\tilde{A}(\bmath{u}_2 - \bmath{u}_2') \tilde{A}(\bmath{u}_3 - \bmath{u}_3')
\delta^2(\bmath{u}_1' +\bmath{u}_2' + \bmath{u}_3' ).
\label{eq:b1}  
\end{equation}
We can divide the region of integration into six separate regions,
$u_1' \geq u_2' \geq u_3'$ and its permutations. Since
$B(u_1', u_2', u_3')$ is invariant under permutations of the $\bmath{u}_i'$,
as is $\delta^2(\bmath{u}_1' + \bmath{u}_2' + \bmath{u}_3' )$,
we can permute the $\bmath{u}_i$ instead of the $\bmath{u}_i'$ to find
\begin{equation}
\langle S(\bmath{u}_1) S(\bmath{u}_2) S(\bmath{u}_3) \rangle =
\int_{u_1' \geq u_2' \geq u_3'} \rmn{d}^2 \bmath{u}_1' \rmn{d}^2 \bmath{u}_2'
\rmn{d}^2 \bmath{u}_3'
B(u_1', u_2', u_3') \sum_{\stackrel{\rmn{perms}}{\{ \bmath{u}_1,
\bmath{u}_2,\bmath{u}_3\}}}
\tilde{A}(\bmath{u}_1 - \bmath{u}_1')
\tilde{A}(\bmath{u}_2 - \bmath{u}_2') \tilde{A}(\bmath{u}_3 - \bmath{u}_3')
\delta^2(\bmath{u}_1' +\bmath{u}_2' + \bmath{u}_3').
\label{eq:b2}
\end{equation}
Transforming to polar coordinates, the delta function restricts the lower
bounds of $u_2'$ and $u_3'$ by the triangle inequality, hence
\begin{eqnarray}
\langle S(\bmath{u}_1) S(\bmath{u}_2) S(\bmath{u}_3) \rangle
&=& \int_0^{\infty} u_1' \rmn{d}u_1' \int_{u_1'/2}^{u_1'} u_2' \rmn{d}u_2'
\int_{u_1' - u_2'}^{u_2'} u_3' \rmn{d}u_3' B(u_1', u_2',u_3')
\nonumber \\ 
&&\mbox{}
\left( \int_0^{2\pi} \rmn{d} \phi_1' \int_0^{2\pi} \rmn{d} \phi_2'
\int_0^{2\pi} \rmn{d} \phi_3'
\sum_{\stackrel{\rmn{perms}}{\{ \bmath{u}_1,
\bmath{u}_2,\bmath{u}_3\}}}
\tilde{A}(\bmath{u}_1 - \bmath{u}_1')
\tilde{A}(\bmath{u}_2 - \bmath{u}_2') \tilde{A}(\bmath{u}_3 - \bmath{u}_3')
\delta^2(\bmath{u}_1' +\bmath{u}_2' + \bmath{u}_3' )\right).
\label{eq:b3}
\end{eqnarray}
For every $u_1'$, $u_2'$ and $u_3'$ within the domain of integration,
the delta function fixes $\phi_2' = \phi_1' + \psi$, where the two
values of $\psi \in [-\pi/2,\pi/2]$ are given by the cosine rule
$2 u_1' u_2' \cos\psi = u_3'{}^2 - u_2'{}^2 - u_1'{}^2$, and $\phi_3'
= \phi_{\bmath{u}_1' + \bmath{u}_2'}+\pi$ closes the triangle. Performing
the (trivial) integrations over $\phi_2'$ and $\phi_3'$, we find that
\begin{eqnarray}
\langle S(\bmath{u}_1) S(\bmath{u}_2) S(\bmath{u}_3) \rangle
&=& \int_0^{\infty} u_1' \rmn{d}u_1' \int_{u_1'/2}^{u_1'} u_2' \rmn{d}u_2'
\int_{u_1' - u_2'}^{u_2'} u_3' \rmn{d}u_3' \frac{2B(u_1', u_2',u_3')}{%
\sqrt{4 u_1'{}^2 u_2'{}^2 - (u_3'{}^2 - u_1'{}^2 - u_2'{}^2)^2}}
\nonumber \\
&&\mbox{}
\left( \int_0^{2\pi} \rmn{d} \phi_1'
\sum_{\stackrel{\rmn{perms}}{\{ \bmath{u}_1,
\bmath{u}_2,\bmath{u}_3\}}}
\sum_{\pm \psi}
\tilde{A}(\bmath{u}_1 - \bmath{u}_1')
\tilde{A}(\bmath{u}_2 - \bmath{u}_2') \tilde{A}(\bmath{u}_3 + \bmath{u}_1'
+ \bmath{u}_2') \right),
\label{eq:b4}
\end{eqnarray}
with $\phi_2'$ fixed to $\phi_1' + \psi$, and the additional summation is
over the two ($\pm$) values of $\psi$. This result is consistent with
equation (C2) of~\citet{Santos} if we transform their result to Fourier space
and replace their weight function $w(\bmath{x})$ by the primary beam of the
interferometer.

To make further progress, we specialise to a Gaussian primary beam. As noted
in Section~\ref{sec:VSA}, this is a good approximation for the VSA
and was assumed for all of the analyses in this paper. For a Gaussian
beam with dispersion $\sigma$, normalised to unity at its peak, the
aperture function is $\tilde{A}(\bmath{u}) = 2\pi \sigma^2 \exp(- 2 \pi^2
\sigma^2 u^2)$. If we write $\bmath{u}_2' = (u_2'/u_1') \textbfss{R}_\psi
\bmath{u}_1'$, where $\textbfss{R}_\psi$ is a right-handed rotation
through $\psi$, then dot products $\bmath{u}_i \cdot \bmath{u}_2'$
can be written as $(u_2'/u_1')(\textbfss{R}_\psi^{-1} \bmath{u}_i)
\cdot \bmath{u}_1'$.
In this manner, the argument of the exponential arising from the product of the
three aperture functions in equation~(\ref{eq:b4}) involves
\begin{equation}
(\bmath{u}_1-\bmath{u}_1')^2 + (\bmath{u}_2-\bmath{u}_2')^2
+ (\bmath{u}_3 + \bmath{u}_1' + \bmath{u}_2')^2 =
\Sigma^2 + \Sigma'{}^2  - 2 \bmath{u}_1' \cdot \bmath{U},
\label{eq:b5}
\end{equation}
where $\Sigma^2 \equiv u_1^2 +u_2^2+ u_3^2$ and
$\Sigma'{}^2 \equiv  u_1'{}^2 +u_2'{}^2+ u_3'{}^2$, and the vector
$\bmath{U} \equiv \bmath{u}_1 - \bmath{u}_3 + (u_2'/u_1')
\textbfss{R}_\psi^{-1} (\bmath{u}_2 - \bmath{u}_3)$. The integration over
$\phi_1'$ can now be performed to give a modified Bessel function $I_0(x)$,
and our result for the three-point function reduces to
\begin{equation}
\langle S(\bmath{u}_1) S(\bmath{u}_2) S(\bmath{u}_3) \rangle
= \! \int_0^{\infty} \!\!\! u_1' \rmn{d}u_1'
\int_{u_1'/2}^{u_1'} \!\!\! u_2' \rmn{d}u_2'
\int_{u_1' - u_2'}^{u_2'} \!\!\! u_3' \rmn{d}u_3' \frac{2(2\pi)^4
\sigma^6 B(u_1', u_2',u_3') e^{-2\pi^2 \sigma^2 (\Sigma^2 + \Sigma'{}^2)}}{%
\sqrt{4 u_1'{}^2 u_2'{}^2 - (u_3'{}^2 - u_1'{}^2 - u_2'{}^2)^2}}
\!\!\!\!\! \sum_{\stackrel{\rmn{perms}}{\{ \bmath{u}_1,
\bmath{u}_2,\bmath{u}_3\}}}
\!\! \sum_{\pm \psi} I_0(4\pi^2 \sigma^2 u_1' U),
\label{eq:b6}
\end{equation}
where $U=|\bmath{U}|$. From this expression, we can easily read off the
complex version of $Q^\alpha_{ijk}$ introduced in Appendix~\ref{sec:appa}.
If we are only interested in triangle configurations,
$\bmath{u}_1 + \bmath{u}_2 + \bmath{u}_3 = 0$, then it is possible to
simplify the product $u_1' U$ further:
\begin{equation}
u_1' U = 3(u_1^2 u_1'{}^2 + u_2^2 u_2'{}^2 + u_3^2 u_3'{}^2)
- \frac{1}{2} \Sigma^2 \Sigma'{}^2 \pm \Delta \Delta', \quad
(\bmath{u}_1 + \bmath{u}_2 + \bmath{u}_3 = 0),
\label{eq:b8}
\end{equation}
where $\Delta \equiv \sqrt{4u_1^2 u_2^2 - (u_3^2 - u_1^2 - u_2^2)}/4$
is the area of the triangle formed from $\bmath{u}_1$, $\bmath{u}_2$ and
$\bmath{u}_3$, and $\Delta'$ is the area of the triangle formed from the primed
vectors. The $\pm$ in equation~(\ref{eq:b8}) arises from the two values of
$\psi$. (Whether $\pm$ corresponds to $\pm\psi$ or $\mp\psi$ depends on the
orientation of the unprimed triangle, but this is irrelevant for the
three-point function since both cases are summed over.)

Finally we note the symmetry properties of the three-point function of the
visibilities. For an azimuthally-symmetric primary beam,
$\langle S(\bmath{u}_1) S(\bmath{u}_2) S(\bmath{u}_3) \rangle$ is invariant
under reflections, rotations and permutations. For triangle configurations,
these symmetries ensure that $\langle S(\bmath{u}_1) S(\bmath{u}_2)
S(\bmath{u}_3) \rangle$ depends only on the lengths $u_1$, $u_2$ and
$u_3$, as is apparent in equation~(\ref{eq:b8}).


\label{lastpage}

\end{document}